\documentclass[aps, prd, superscriptaddress, nofootinbib, preprintnumbers,twocolumn]{revtex4}

\usepackage{amsmath,amssymb,bm}
\usepackage[dvipdfmx]{graphicx}
\usepackage{graphicx,bm,color}
\usepackage{float}
\usepackage{braket}
\usepackage{subfigure}

\newcommand{\cred}[1]{{\color{black}#1}} 

\newcommand{\cblue}[1]{{\color{black}#1}} 

\newcommand{\ccred}[1]{{\color{black}#1}}

\allowdisplaybreaks

\begin{document}

\title{
Chiral-Imbalance Density Wave in Baryonic Matters  
}

\author{Mamiya Kawaguchi\footnote{mkawaguchi@hken.phys.nagoya-u.ac.jp}}
      \affiliation{ Department of Physics, Nagoya University, Nagoya 464-8602, Japan.} 
\author{Shinya Matsuzaki\footnote{synya@jlu.edu.cn}}   
      \affiliation{Center for Theoretical Physics and College of Physics, Jilin University, Changchun, 130012, China}  

\date{\today}

\def\theequation{\thesection.\arabic{equation}}
\makeatother

 \begin{abstract}
We propose a new chirality-imbalance phenomenon arising 
in baryonic/high dense matters under a magnetic field.  
A locally chiral-imbalanced (parity-odd) domain can be created 
due to the electromagnetically induced $U(1)_A$ anomaly in  
high-dense matters.  
The proposed local-chiral imbalance generically possesses a close relationship 
to a spacial distribution of an inhomogeneous chiral (pion)-vector current coupled to the magnetic field. 
To demonstrate such a nontrivial correlation, we take 
the skyrmion crystal approach to model baryonic/high dense matters.  
Remarkably enough,  
we find the chirality-imbalance distribution takes a wave form  
in a high density region (dobbed ``{\it chiral-imbalance density wave}''), 
when the inhomogeneous chiral condensate 
develops to form a chiral density wave. 
This implies the emergence of 
a nontrivial density wave for the explicitly broken $U(1)_A$ current 
simultaneously with the chiral 
density wave for the spontaneously broken chiral-flavor current. 
We further find that 
the topological phase transition in the skyrmion crystal model 
(between skyrmion and half-skyrmion phases) undergoes   
the deformation of the chiral-imbalance density wave in shape 
and periodicity. 
The emergence of this chiral-imbalance density wave 
could give a crucial contribution 
to studies on the chiral phase transition,  
as well as the nuclear matter structure, 
in compact stars under a magnetic field. 

\end{abstract}
\maketitle
\section{Introduction}
Exploring the properties of QCD under an extreme
environment has attracted a lot of attentions 
to extract a novel insight of the nonperturbative nature of QCD 
involving the chiral-symmetry breaking structure.   
Particularly, it would be important to ask how much 
the $U(1)_A$ breaking (anomaly) can serve as a source for 
the baryonic matter structure, while competing with contributions from 
the spontaneous breaking of the chiral symmetry,  
which would also be linked to 
the origin of the \cred{nucleon} mass.

Regarding a nontrivial issue on the $U(1)_A$ anomaly under 
exotic environments,  
\cred{it has been expected~\cite{Kharzeev:2001ev,Kharzeev:2007tn,Kharzeev:2007jp,Fukushima:2008xe,Son:2009tf,Kharzeev:2010gd,Burnier:2011bf,Hongo:2013cqa,Yee:2013cya,Hirono:2014oda,Adamczyk:2015eqo,Yin:2015fca,Huang:2015oca,Kharzeev:2015znc,
Hattori:2016emy,Shi:2017cpu}  
in the hot QCD system} that 
the local-parity-odd domain 
would show up due to the nonzero chirality (sometimes called the chiral-charge separation) induced by the anomalous 
$U(1)_A$ current coupled to the topological gluon configuration 
(sphaleron~\cite{Manton:1983nd,Klinkhamer:1984di}). 
This local $P$-odd domain, called the chiral-imbalance medium~\cite{McLerran:1990de},  
is expected to be observed as the nontrivial consequences 
for the presence of the $U(1)_A$ anomaly (and/or strong CP violation) 
in heavy ion collisions, 
although being metastable to be gone after the typical time scale of QCD pasts~\cite{McLerran:1990de,Moore:1997im,Moore:1999fs,Bodeker:1999gx,deSouza:2015ena}.  
Such a chiral imbalance medium 
is characterized by the chiral (axial) chemical potential 
(often denoted by $\mu_5$), and has so far extensively been searched based on 
various arguments in hot QCD matter 
applied to heavy ion collision experiments, \cred{e.g.~\cite{Kharzeev:2001ev,Kharzeev:2007tn,Kharzeev:2007jp,Fukushima:2008xe,Son:2009tf,Kharzeev:2010gd,Burnier:2011bf,Hongo:2013cqa,Yee:2013cya,Hirono:2014oda,Adamczyk:2015eqo,Yin:2015fca,Huang:2015oca,Kharzeev:2015znc,Hattori:2016emy,Shi:2017cpu}   
and} 
\cite{Andrianov:2012hq,Andrianov:2012dj,Kawaguchi:2016avk,Andrianov:2017hbf,Andrianov:2017ilv,Andrianov:2018gim,Putilova:2018qli,Andrianov:2019you}.

In place of invoking the gluonic $U(1)_A$ anomaly, 
a local chiral-imbalance medium can actually be created  
even in other specific environments apart from QCD.  
For instance, weak interactions generically \cred{break} the parity and \cred{lead} to 
the chirality imbalance. 
\cred{Indeed, it has recently been shown~\cite{Charbonneau:2009ax,Grabowska:2014efa,Kaminski:2014jda,Sigl:2015xva,Yamamoto:2015gzz,Masada:2018swb} that} 
the chirality imbalance for weakly interacting leptons can be generated 
in a process of supernova explosions, \cred{through the chiral transport mechanism 
acting on neutrinos}.  
Moreover, a signal induced from the chiral imbalance medium (so-called the chiral 
magnetic effect) has actually been observed in a condense matter system, called Weyl semimetals~\cite{Li:2014bha,Lv:2015pya,Xu:2015cga}. 
Thus, understanding of the chirality imbalance as well as the $U(1)_A$ anomaly 
has recently \cred{been developing} and is currently getting interdisciplinary.

In this paper, we propose a novel 
possibility to create a (stable) chiral-imbalance 
medium in a high dense QCD with a strong magnetic field. 
Our central idea is built on a simple observation as follows.

In zero-temperature environment, 
the topologically nontrivial gluon configuration will not survive longer enough 
to stay in the QCD time scale, due to the gigantic exponential suppression form 
of the instanton configuration. 
Thereby, the $U(1)_A$ anomaly of the 
gluon-field strength form $\sim \epsilon_{\mu\nu\rho \sigma }G^{\mu\nu} G^{\rho \sigma}$   
is supposed to undetectable, in contrast to the finite temperature 
case with the QCD sphaleron configuration. 
Instead of the gluonic contribution, 
the chiral imbalance for quarks 
can be induced by the electromagnetic $U(1)_A$ anomaly, 
where the $U(1)_A$ symmetry is explicitly broken by coupling the chiral quarks 
to the electromagnetic current, yielding the anomaly form like 
$\epsilon_{\mu\nu\rho \sigma }F^{\mu\nu} F^{\rho \sigma}$, 
where $F_{\mu\nu}$ stands for the electromagnetic filed strength.

Now, consider a high-dense matter system under a magnetic field,  
inside of which charged pions form the pion-vector current 
$J_\mu^\pi =  i (\partial_\mu \pi^+ \pi^- - \pi^+ \partial_\mu \pi^-)$, 
coupled to the electromagnetic field, and then might also couple 
to the electromagnetic $U(1)_A$ anomaly as above. 
Of importance here is to note that 
as discussed in~\cite{Nishiyama:2015fba,Abuki:2016zpv,Abuki:2018wuv,Kawaguchi:2018xug}, 
in a high-dense medium with a strong magnetic field, 
the pions can locally form condensates, 
(so-called inhomogeneous chiral condensates), 
so that the pion-vector current $J_\mu^\pi$ 
can also have a locally nontrivial distribution in the medium. 
Suppose the medium to be so high-dense like that, in such a way that 
the dense matter can be highly compressed to be almost static. 
In that case, we may expect to have a local-chiral density 
($\rho_5(\vec{x})$ with $\vec{x}$ being three-dimensional spatial vector)  
associated with the $U(1)_A$ anomaly in the high-dense medium, like  
\begin{align}
 \rho_5(\vec{x}) 
& \sim   \epsilon^{0 \mu\nu\rho} J_\mu^\pi(\vec{x}) F_{\nu\rho}(\vec{x})/f_\pi^2 
 \notag \\ 
&\sim \vec{J}^\pi(\vec{x}) \cdot \vec{B}(\vec{x})/f_\pi^2  
\,. \label{generic-rho5}
\end{align}

The nonzero $\rho_5$ in Eq.(\ref{generic-rho5}) 
manifestly implies the emergence of a chirality-imbalance driven 
in the presence of a magnetic field and a nontrivial spatial distribution for 
a pion-vector current. 
In fact, \cred{one could immediately find the existence of the interaction terms as in Eq.(\ref{generic-rho5}), once writing down the conventional chiral Lagrangian including 
the chiral anomaly part:}  
the interaction terms of this type can be generated 
via the Wess-Zumino-Witten (WZW) term~\cite{Wess:1971yu,Witten:1983tw}, 
when the $U(1)_A$ charge 
and the isospin charge plus baryon number are gauged to be 
identified with the chiral chemical potential $\mu_5$ and the electromagnetic 
charge for quarks, respectively. 
Then the functional derivative of the chiral Lagrangian 
with respect to the $\mu_5$ will immediately lead to 
the chiral-imbalance density in Eq.(\ref{generic-rho5})~\footnote{
Several discussions on chirality-imbalance effects,  
induced with a strong magnetic field for hadron physics,  
arising through the WZW term, have been made 
\cred{so far~\cite{Fukushima:2012fg,Cao:2015cka}}. 
However, to our best knowledge, 
no references have picked up the interaction term as in Eq.(\ref{generic-rho5}), 
which involves the pion-vector current part, instead of the external photon 
field.}. 
Therefore, might it sound trivial? --- the answer is ``No.''.  
\cred{From} Eq.(\ref{generic-rho5}) 
one would expect a nontrivial correlation between
the presence of the chirality imbalance and the inhomogeneity of 
the pion-vector current (arising from the inhomogeneous 
chiral condensate) in medium. 
That is, Eq.(\ref{generic-rho5}) would provide a significant possibility 
to examine {\it how much the $U(1)_A$ anomaly can be correlated with 
the spontaneously broken chiral symmetry, by setting the chiral dynamics 
in a high-dense medium with use of a magnetic field as the probe}.

To monitor a nontrivial physics derived from the $\rho_5$ in Eq.(\ref{generic-rho5}), 
in this paper we take the skyrmion crystal approach~\cite{Skyrme:1962vh,Klebanov:1985qi} 
as a candidate effective model for describing the baryonic/high-dense matter. 
In the skyrmion crystal approach, 
the baryonic matter is described by putting the skyrmions on lattice vertices of a crystal structure. 
Actually, the large-$N_c$ QCD supports that
a topologically-static soliton arising as a skyrmion can be regarded as a baryon~\cite{Witten:1983tw,Witten:1983tx}. 
By using the skyrmion crystal approach,
the baryonic matter can also be described as the topologically static-object. 
In high density region, where baryons is so compressed to be a static object, 
the skyrmion crystal is a powerful approach for the 
\ccred{qualitative-baryon description}~\footnote{
\ccred{Going beyond such qualitative arguments on the baryon description, 
it has recently been suggested that the skyrmion approach could quantitatively 
be consistent with realistic light nuclei with a desirable size of the binding 
energy~\cite{Naya:2018kyi}. 
}}
as if they could form crystals~\cite{Lee:2003aq,Lee:2003rj,Ma:2013ooa,Ma:2013ela,book,Ma:2016npf,Ma:2016gdd,Harada:2015lma,Kawaguchi:2018xug,Kawaguchi:2018fpi}.

The skyrmion crystal picture 
is indeed in accord with the desired setup for the emergence of 
a local-chirality imbalance proposed in Eq.(\ref{generic-rho5}).   
Besides, the skyrmion crystal approach predicts a characteristic phenomena which
is called “topological phase transition"
\cblue{(for reviews see e.g., Refs.~\cite{book,Ma:2016npf,Ma:2016gdd})}. 
If we choose the underlying structure as the face-centered-cubic (FCC) crystal,
the crystal configuration is changed from a FCC crystal 
to a cubic-centered crystal (CC). 
\cblue{
Actually,
\ccred{it has been indicated} that 
the results from effective field theories, in which such a topological phase transition
is encoded,   
\ccred{can be} consistent with the present observation of neutron star 
physics~\cite{Paeng:2017qvp,Ma:2018jze,Ma:2018xjw}. 
} 
\ccred{Those facts} would give us another interesting chance to investigate 
some correlations between the chirality imbalance and the baryonic 
matter structure.

We demonstrate that 
a nontrivial correlation between the chiral imbalance distribution and 
the baryonic/high-dense matter structure actually shows up:  
it turns out 
that the $\rho_5$ as in Eq.(\ref{generic-rho5})  
emerges to form a density wave, simultaneously with a chiral 
density wave for a pion-vector current $J^\pi_\mu$. 
It is dobbed ``{\it chiral-imbalance density wave}''. 
We also observe that 
the periodicity of the chiral-imbalance density wave 
harmonizes to almost coincide with the previously proposed 
inhomogeneous-chiral condensate distributions induced on the skyrmion 
crystal~\cite{Kawaguchi:2018xug}.

The emergence of this chiral-imbalance density wave 
would significantly contribute  
to studies on the chiral phase transition under a magnetic field 
with the inhomogeneous chiral condensates (chiral density waves) 
incorporated, as has been discussed in~\cite{Nishiyama:2015fba,Abuki:2016zpv,Abuki:2018wuv} 
in different setup for chiral effective models, 
and also on the nuclear matter structure, 
in compact stars holding a magnetic field.

This paper is organized as follows: 
In Sec. II we start with introduction of our target model-setup, by  
reviewing the skyrmion crystal approach  
for modeling of baryonic matters. 
This part also includes introduction of a magnetic field and 
a chiral chemical potential in the chiral effective model, 
so as to examine the chirality imbalance 
induced by a magnetic field. 
Then we explicitly 
see that the chiral imbalance density as proposed in Eq.(\ref{generic-rho5}) 
indeed shows up in the present setup. 
Quantities related to the inhomogeneous 
chiral condensate as well 
as the baryon number density are also derived there. 
Sec. III provides the numerical analysis on the chiral imbalance distributions 
in the skyrmion crystal, and shows the emergence of the chira-imbalance 
density wave, and several related phenomena, such as correlation 
with the inhomogeneity of the chiral condensate.  
\cred{Conclusion} for the present paper is given in Sec.IV. 
Appendix A compensates knowledge on 
symmetry properties for the chiral-imbalance 
distribution and the inhomogeneous chiral condensate, 
in the skyrmion crystal.

\section{Chirality imbalance in skyrmion crystal under a magnetic field}
In this section,
we first introduce preliminary setups in studying the magnetic properties of the skyrmion crystal, such as the basic construction in the chiral limit (Part A)
and see how the baryon number density is modified by the presence of a 
magnetic field which explicitly breaks the chiral symmetry (Part B).   
And then, we discuss how the magnetic-field driven $U(1)_A$ anomaly 
induces the chirality imbalance (Part C).

\subsection{Skyrmion crystal approach}

The Skyrme model \cite{Skyrme:1962vh} based on 
the 2-flavor chiral symmetry is described in the chiral limit 
by the following Lagrangian: 
\begin{eqnarray}
{\cal L}_{\rm Skyr} & = & \frac{f_\pi^2}{4}{\rm tr}\left[\partial_\mu U \partial^\mu U^\dagger\right] + \frac{1}{32g^2}{\rm tr}\left[U^\dagger \partial_\mu U,U^\dagger \partial_\nu U\right]^2,
\nonumber\\
\label{Lag1}
\end{eqnarray}
where $U$ is the chiral field parameterized by the pion fields, $ f_\pi$ the pion decay constant, and $g$ the dimensionless coupling constant. 
In the skyrmion crystal approach, the chiral field $U$ is parameterized as
\begin{eqnarray}
U=\phi_0+i\tau_a\phi_a,
\label{paraU}
\end{eqnarray}
with $a = 1,2,3$, $\tau^a$ being the Pauli matrices,  
 and the unitary constraint $(\phi_0)^2+(\phi_a)^2=1$. 
For later convenience, we also introduce unnormalized fields $\bar \phi_\alpha\;(\alpha=0,1,2,3)$, which are related to the corresponding normalized ones through
\begin{eqnarray}
\phi_\alpha=
\frac{\bar\phi_\alpha}{\sqrt{\sum_{\beta=0}^3\bar\phi_\beta\bar\phi_\beta}}.
\end{eqnarray}

We will consider the static skyrmion crystal formed by the static pion fields, $\phi_\alpha(t,x,y,z)=\phi_\alpha(x,y,z)$.
In a crystal lattice with a periodicity of $2L$ (the size of the unit cell for a single crystal),
the static pion fields can be expanded in terms of the Fourier series~\cite{Lee:2003aq}:
\begin{align}
\bar \phi_0(x,y,z)&=\sum_{a,b,c}\bar \beta_{abc}\cos(a\pi x/L)\cos(b\pi y/L)\cos(c\pi z/L)\nonumber\\
\bar \phi_1(x,y,z)&=\sum_{h,k,l}\bar \alpha_{hkl}^{(1)}\sin(h\pi x/L)\cos(k\pi y/L)\cos(l\pi z/L)\nonumber\\
\bar \phi_2(x,y,z)&=\sum_{h,k,l}\bar \alpha_{hkl}^{(2)}\cos(l\pi x/L)\sin(h\pi y/L)\cos(k\pi z/L)\nonumber\\
\bar \phi_3(x,y,z)&=\sum_{h,k,l}\bar \alpha_{hkl}^{(3)}\cos(k\pi x/L)\cos(l\pi y/L)\sin(h\pi z/L)\,, 
\label{ansatz_1}
\end{align}
where $h, k$ and $l$ are taken to be integers. 
In the present study, we shall construct the FCC crystal from the skyrmion approach.  
Then we need to impose some constraint conditions on
the Fourier coefficient $\bar{\alpha}$ and $\bar{\beta}$ (for more on this, see~\cite{Lee:2003aq}).

It is also interesting to note that in the skyrmion crystal approach
the configuration of the $\phi_\alpha(x,y,z)$ can be rephrased as 
the inhomogeneous quark condensates like, 
\begin{eqnarray}
\phi_0(x,y,z) &\sim& \langle 0|\bar q q |0\rangle(x,y,z)\nonumber \\
\phi_{a} (x,y,z) &\sim&
\langle 0|\bar q i\gamma_5\tau_{a} q |0\rangle(x,y,z).
\label{qcond}
\end{eqnarray}

The skyrmion crystal approach has the characteristic phenomena which is so-called
topological phase transition.
At some critical lattice size, this phase transition occurs in the skyrmion crystal.
As a result, the FCC crystal structure changes to be 
the CC form.
In terms of the phase transition, 
the low density region with the skyrmion crystal
realized as the FCC form is called “skyrmion phase", while the high density
region where the CC structure is manifest, called “half-skyrmion phase"~\cite{Lee:2003aq,Lee:2003rj,Ma:2013ooa,Ma:2013ela,book,Ma:2016npf,Ma:2016gdd,Harada:2015lma,Kawaguchi:2018xug,Kawaguchi:2018fpi}.

\subsection{Baryon number density in a magnetic field} 
To consider the magnetic effect on the skyrmion crystal,
we introduce the external vector field ${\cal V}_\mu$ 
and the external axial field ${\cal A}_\mu$, by gauging the chiral symmetry, 
\begin{eqnarray}
D_\mu U=\partial_\mu U-i[{\cal V}_\mu, U]+i\{{\cal A}_\mu,U\}.
\end{eqnarray}
The magnetic field ($B$), the baryon chemical potential ($\mu_B$) and the chiral chemical potential ($\mu_5$)
are embedded into the ${\cal V}_\mu$ and ${\cal A}_\mu$,
\begin{eqnarray}
{\cal V}_\mu&=& Q_B\, \mu_B \delta_{\mu0}
+eQ_{\rm em} A_\mu\nonumber\\
{\cal A}_\mu&=&
\mu_5 \delta_{\mu0}
{\bm 1}_{2\times 2},
\end{eqnarray}
where
$Q_{B}=\frac{1}{3}{\bm 1}_{2\times 2}$ is the baryon number charge matrix,
$Q_{\rm em}=\frac{1}{6}{\bm 1}_{2\times 2}+\frac{1}{2}\tau_3$ the electric charge matrix,  
and 
the magnetic field $B$ is incorporated into $A_\mu$.
In this study, 
we consider a constant magnetic field $(B)$ along the z axis. 
Then, the magnetic field breaks the $O(3)$ symmetry down to the  $O(2)$ symmetry.
To respect the residual $O(2)$ symmetry,
we choose the following symmetric gauge,
\begin{eqnarray}
A_\mu=
-\frac{1}{2}B y\delta_\mu^{\;\;1}+\frac{1}{2}Bx \delta^{\;\;2}_{\mu}.
\label{symgauge}
\end{eqnarray}


The covariantized 
Wess-Zumino-Witten (WZW) action ($\Gamma_{\rm WZW}=\int d^4x {\cal L}_{\rm WZW}$),   
which corresponds to the solution for the 
non-Abelian $U(2)_L \times U(2)_R$ anomaly equation, 
makes the baryon number density $\rho_B$ 
coupled to the baryon chemical potential $\mu_B$. 
Hence, the $\rho_B$ is given by~\cite{Son:2004tq,Son:2007ny,He:2015zca} 
\begin{eqnarray}
 \rho_B&=&\frac{\partial{\cal L}_{\rm wzw}}{\partial \mu_B}\Biggl|_{\mu_B=0}
=
 \rho_W+\tilde\rho_{eB}
 \end{eqnarray}
 with
\begin{align}
 \rho_W&= 
 \frac{1}{24\pi^2}\epsilon^{0\nu\rho\sigma}{\rm tr}
\left[
(\partial_\nu U\cdot U^\dagger)(\partial_\rho U\cdot U^\dagger)(\partial_\sigma U\cdot U^\dagger)
\right]
 \notag\\
\tilde\rho_{eB} 
&= 
\frac{i e}{16\pi^2}\epsilon^{0\nu\rho\sigma} 
\partial_\sigma \left( 
{\rm tr} [ 
A_\nu Q_{\rm em}
\{ \partial_\rho U, U^\dag  \}
]  
\right) 
\,, 
\label{rhoeB}
\end{align}
where $\rho_W$ denotes the winding-number density and
$\tilde\rho_{eB}$ is the induced baryon number density.
By taking the symmetric gauge in Eq.(\ref{symgauge}),
$ \rho_B$ 
is evaluated as a function of a set of the topological $\phi_\alpha$ fields (Eq.(\ref{paraU})) 
as follows: 
\begin{align}
\rho_B&=\frac{1}{24\pi^2}\epsilon^{0\nu\rho\sigma}{\rm tr}
\left[
(\partial_\nu U\cdot U^\dagger)(\partial_\rho U\cdot U^\dagger)(\partial_\sigma U\cdot U^\dagger)
\right] \Bigg|_{\phi_{\alpha}}
 \nonumber\\
 &
 - \frac{eB}{4\pi^2}\left[(\partial_z\phi_3)\phi_0-(\partial_z\phi_0)\phi_3 \right]\nonumber\\
 &
 + \frac{eB}{8\pi^2} \bigl(\left\{
[y\partial_y\phi_3]_{\rm disc}(\partial_z\phi_0)-[y\partial_y\phi_0]_{\rm disc}(\partial_z\phi_3)\right\}  
 \notag \\ 
&  -
\left\{
(\partial_z\phi_3)[x\partial_x\phi_0]_{\rm disc}-(\partial_z\phi_0)[x\partial_x\phi_3]_{\rm disc}
\right\}\bigl)
\,. 
\label{discsymmetricBD}
\end{align}
Here 
we have used discretized form with a derivative operator such as $[y\partial_y\phi_3]_{\rm disc}$
to hold the translational invariance in the skyrmion crystal
(the explicit expressions and more details are supplied in \cite{Kawaguchi:2018xug}).

In the skyrmion crystal approach, the skyrmion crystal configuration
can be visualized through the baryon-number density distribution $\rho_B(x,y,z)$.
Therefore the external magnetic field deforms the skyrmion configuration  
due to the existence of the induced baryon number density, as discussed in \cite{Kawaguchi:2018xug}.

\subsection{Induced chiral-imbalance density}
Now we discuss how the chirality imbalance shows up  
under a magnetic field through the $U(1)_A$ anomaly. 
We first note that 
in a way similar to the baryon number density, 
the covariantized WZW action regarding the $U(2)_L \times U(2)_R$ anomaly 
makes it possible to couple the chirality imbalance functional $\rho_5$ (hereafter called chiral imbalance distribution) with the chiral chemical potential $\mu_5$.  
It arises from the ${\cal V}-{\cal V}-{\cal A}$ 
type interaction terms, to be  
\begin{align}
\rho_5&=\
\frac{\partial{\cal L}_{\rm wzw}}{\partial \mu_5}\Biggl|_{\mu_5=0}
=
 \frac{ie }{16\pi^2}\epsilon^{0\nu\rho\sigma} 
\partial_\nu \left( {\rm tr} [A_\rho Q_{\rm em} [\partial_\sigma U, U^\dag]] \right) 
\notag \\ 
&= 
\frac{e }{8 \pi^2} 
{\rm tr} \left[ Q_{\rm em} \left\{ 
\tilde{F}^{0\mu} H_\mu(U) + \tilde{F}^{0\mu}(U) A_\mu \right\}  \right] 
\,, 
\label{rho5:rewritten}
 \end{align}
where $H_\mu(U) = i [\partial_\mu U, U^\dag]$ denotes a pion-vector current field 
(sometimes called a chiral connection field) and $\tilde{F}^{0\mu} \equiv \frac{1}{2} \epsilon^{0\mu\nu\rho} F_{\nu \rho}$ is a generic magnetic field strength.  
Thus the chiral imbalance density $\rho_5$ is 
supplied by the pionic vector current coupled to the external magnetic field.

At this moment, we have arrived at our central formula, Eq.(\ref{rho5:rewritten}),  
which is a generalization form for Eq.(\ref{generic-rho5}). 
Indeed, this is in accord with the form of the $U(1)_A$ anomaly $\sim \tilde{F}_{\mu\nu} F^{\mu\nu}$, 
where one of the field strength $F_{\mu\nu}$ is replaced by the one 
constructed from \cred{the pionic-vector current $H_\mu(U)$}.

Projecting the generic formula of Eq.(\ref{rho5:rewritten}) onto 
the skyrmion crystal approach, 
we see that 
the $\rho_5$ is expressed by a set of skyrmion configurations parametrized by 
the topological $\phi_\alpha$ fields, as in Eq.(\ref{paraU}). 
Under the symmetric gauge in Eq.(\ref{symgauge}),
the chiral imbalance distribution $\rho_5$ is thus evaluated as
\begin{align} 
\rho_5&=
\frac{eB}{4\pi^2}\bigl\{(\partial_z\phi_1)\phi_2-(\partial_z\phi_2)\phi_1\bigl\} 
\notag\\
&
+ \frac{eB}{8\pi^2}\bigl(
[x\partial_x\phi_1]_{\rm disc}
+[y\partial_y\phi_1]_{\rm disc}
\bigl)(\partial_z\phi_2)
\notag \\ 
& -
\frac{ eB}{8\pi^2}\bigl(
[x\partial_x\phi_2]_{\rm disc}
+[y\partial_y\phi_2]_{\rm disc}
\bigl)(\partial_z\phi_1)
\,. 
\label{symmetricCID}
\end{align}
Here 
we have used the discretized form in similar way to Eq.(\ref{discsymmetricBD}).
Crucial to note here is that 
no chirality imbalance distribution emerges without 
nonzero magnetic field. 
Note also that
the space averaged value of $\rho_5$ goes to zero:
$\int_{\rm cube} d^3 x \rho_5=0$. 
This is because of the parity-odd property. 
(Or, more generically, it is due to the fact that 
no nontrivial configuration is presented  
for the external gauge field $A_\mu$ at the boundary of the target cube. 
See Eq.(\ref{rho5:rewritten}).) 
Hence, 
these facts imply that 
the skyrmion crystal is turned  
into the local-chiral imbalance medium
by the presence of a magnetic field, 
in which the local-chiral imbalance distribution 
would be expected to have a nontrivial 
correlation with the local-inhomogeneous chiral condensates in Eq.(\ref{qcond}), 
as well as the local-baryon number density distribution given by $\rho_B$ 
in Eq.(\ref{discsymmetricBD}).

\section{Numerical results}
In this section 
we numerically examine the chiral imbalance distribution $\rho_5$ in Eq.(\ref{symmetricCID})
and make an attempt to find its nontrivial correlation with the baryon matter structure,  
based on the skyrmion crystal approach under a magnetic field. 
The baryon matter structure can be monitored by 
examining the position dependence  
of $\rho_B$  
in Eq. (\ref{discsymmetricBD}), and the chiral imbalance distribution $\rho_5$ by  
Eq.(\ref{symmetricCID}).  
We then note that 
the baryon number density $\rho_B$ and the chiral imbalance distribution $\rho_5$ 
are expressed as the function of the Fourier coefficients $\bar\beta_{abc}$ and 
$\bar\alpha_{hkl}^{(i)}$, 
as seen from Eq.~(\ref{ansatz_1}). 
There the Fourier coefficients 
are determined by minimizing the per-baryon energy 
$ 
E/N_ B=-
\frac{1}{4}\int_{\rm cube}d^3x {\cal L}_{\rm Skyr}, 
$ 
with $N_B$ having been taken to be 4, and 
$\int_{\rm cube}=\int_{-L}^{L}dx\int_{-L}^{L}dy\int_{-L}^{L}dz$.
(Note that the magnetically induced $\tilde{\rho}_{\rm eB}$ 
in Eq.(\ref{rhoeB}) 
vanishes in the integral, because of the trivial configuration 
for the gauge field $A_\mu$ at the boundary of the target cube, 
hence it does not affect the total baryon number at all.)  
Then, 
 the per-baryon energy is also expressed as a function of the Fourier coefficients $\bar\beta_{abc},\bar\alpha_{hkl}^{(i)}$ 
 which are used as variational parameters in the numerical calculation.
Once the strength of a magnetic field is fixed, 
the Fourier coefficients for a given set of crystal size $L$ is determined  
by minimizing the per-baryon energy.
Thus,
the Fourier coefficients $\bar\beta_{abc},\bar\alpha_{hkl}^{(i)}$ depend on the crystal size $L$ and a magnetic field scale $eB$. 
For numerical computations, 
we take $f_\pi=92.4\,{\rm MeV}$ and $g=5.93$ 
as inputs~\cite{Ma:2016npf}.

\subsection{Chiral imbalance distribution on the skyrmion crystal: 
chiral-imbalance density wave}
In this subsection we examine the \cred{chirality imbalance} on the skyrmion crystal configuration in the presence of a magnetic field,
which 
can be visualized through the chiral imbalance distribution, $\rho_5$, and
the baryon-number density distribution, $\rho_B$, respectively.

In Figs.~\ref{eB400L200} and~\ref{eB800L200},
we plot the skyrmion crystal configurations and the chiral imbalance distribution $\rho_5$ in a low density region where $L=2.0\,{\rm fm}$ 
(corresponding to the skyrmion phase), with 
the magnetic field scale fixed to 400 MeV and 800 MeV, respectively. 
\cred{The magnitude of $\rho_5$ has been amplified by multiplying a factor of 10, 
because of its smallness compared to the baryon number density $\rho_B$}.    

First, see the left panels in the figures (the panels (a)),  
showing the skyrmion crystal configurations characterized by the baryon number density 
$\rho_B$ under a magnetic field. 
We then find that even in the presence of a magnetic field,
the skyrmion crystal keeps the FCC structure, as was discussed in \cite{Kawaguchi:2018xug}. 

Looking at the middle panels (b) in  
Figs. \ref{eB400L200} and \ref{eB800L200}: 
one realizes that remarkably nontrivial phenomenon has been 
emergent.  
These panels display 
the chiral imbalance distributions in the skyrmion crystal, 
and show that  
the  chirality imbalance shows up 
on the skyrmion crystal to be locally distributed on the FCC crystal configuration. 
Interesting enough, such a chiral imbalance distribution 
looks like forming a wave (with odd parity). 
This can be dubbed  
``{\it chiral-imbalance density wave}'', which flows quite differently from  
the baryon number density on the crystal.

From the left panels in the figures (panels (c)),  
we also see that 
the magnitude of the chiral imbalance distribution gets bigger,
as a magnetic field increases, which is 
simply expected from the form of the functional of $\rho_5$ 
in Eq.(\ref{symmetricCID}).  
 

Next, in Figs. \ref{eB400L100} and  \ref{eB800L100}, 
we draw the skyrmion configurations and 
the chiral imbalance distribution in a high density region 
where $L=1.0\,{\rm fm}$
(corresponding to the half-skyrmion phase), with $\sqrt{eB}=400$ and $800$ MeV, 
respectively. 
At the first glance,
one immediately finds that 
as a magnetic field gets bigger, 
the CC configuration dramatically becomes distorted
(Figs. \ref{eB400L100}(a) and  \ref{eB800L100}(a)), as was observed 
in \cite{Kawaguchi:2018xug}. 

As for the chiral imbalance distribution, 
again, the configuration of $\rho_5$ forms a wave (the chiral-imbalance density wave), 
which, however, looks   
quite similar to the skyrmion configuration 
except for the parity property. 
This is in contrast to the case of the skyrmion phase, 
as depicted in Figs. \ref{eB400L100}(b) and  \ref{eB800L100}(b). 
This observation indicates a nontrivial consequence 
that the topological phase transition leads to  
the change of the chiral-imbalance density wave in shape. 

Similarly to the case for the skrymion phase, 
Figs. \ref{eB400L100}(c) and  \ref{eB800L100}(c) also show that
the chiral imbalance distribution grows up as the baryonic matter approaches 
a higher-intense object influenced by 
a strong magnetic field (with $\sqrt{eB}=800\,{\rm MeV}$). 


\begin{widetext}

\begin{figure}[H]
\begin{tabular}{cc}
 \begin{minipage}{0.31\hsize}
  \begin{center}
   \includegraphics[width=5.5cm]{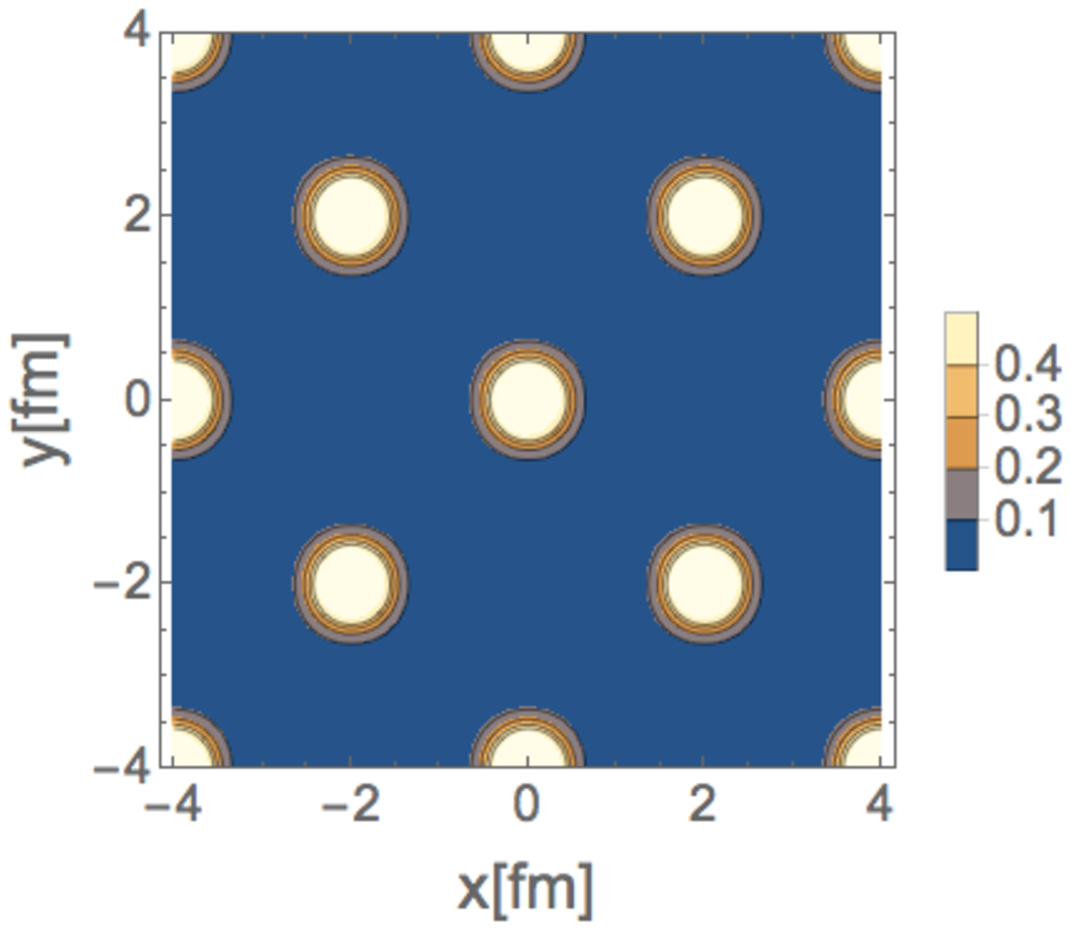}
    \subfigure{(a)}
  \end{center}
 \end{minipage} 
 \begin{minipage}{0.31\hsize}
  \begin{center}
   \includegraphics[width=5.5cm]{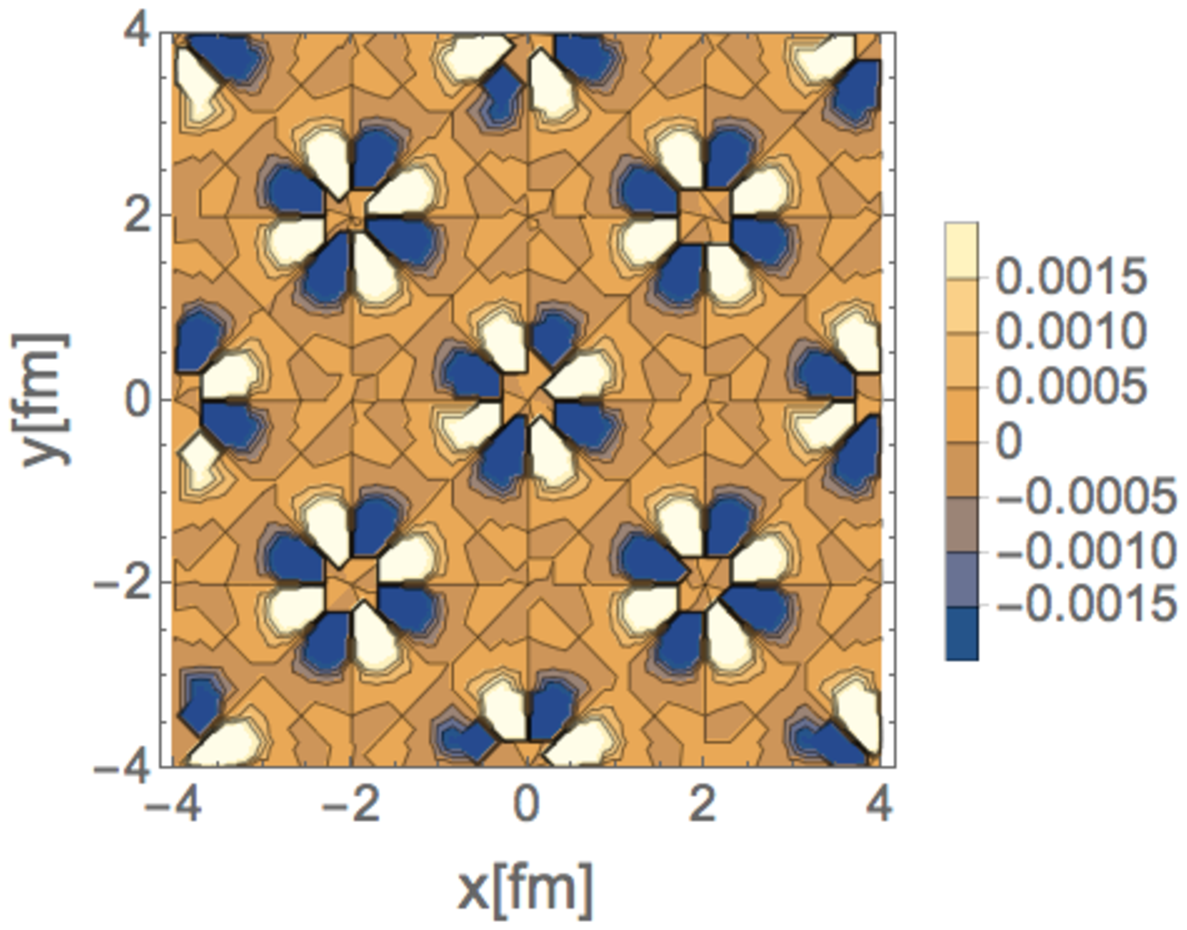}
    \subfigure{(b)}
  \end{center}
 \end{minipage}
  \begin{minipage}{0.31\hsize}
  \begin{center}
   \includegraphics[width=5.7cm]{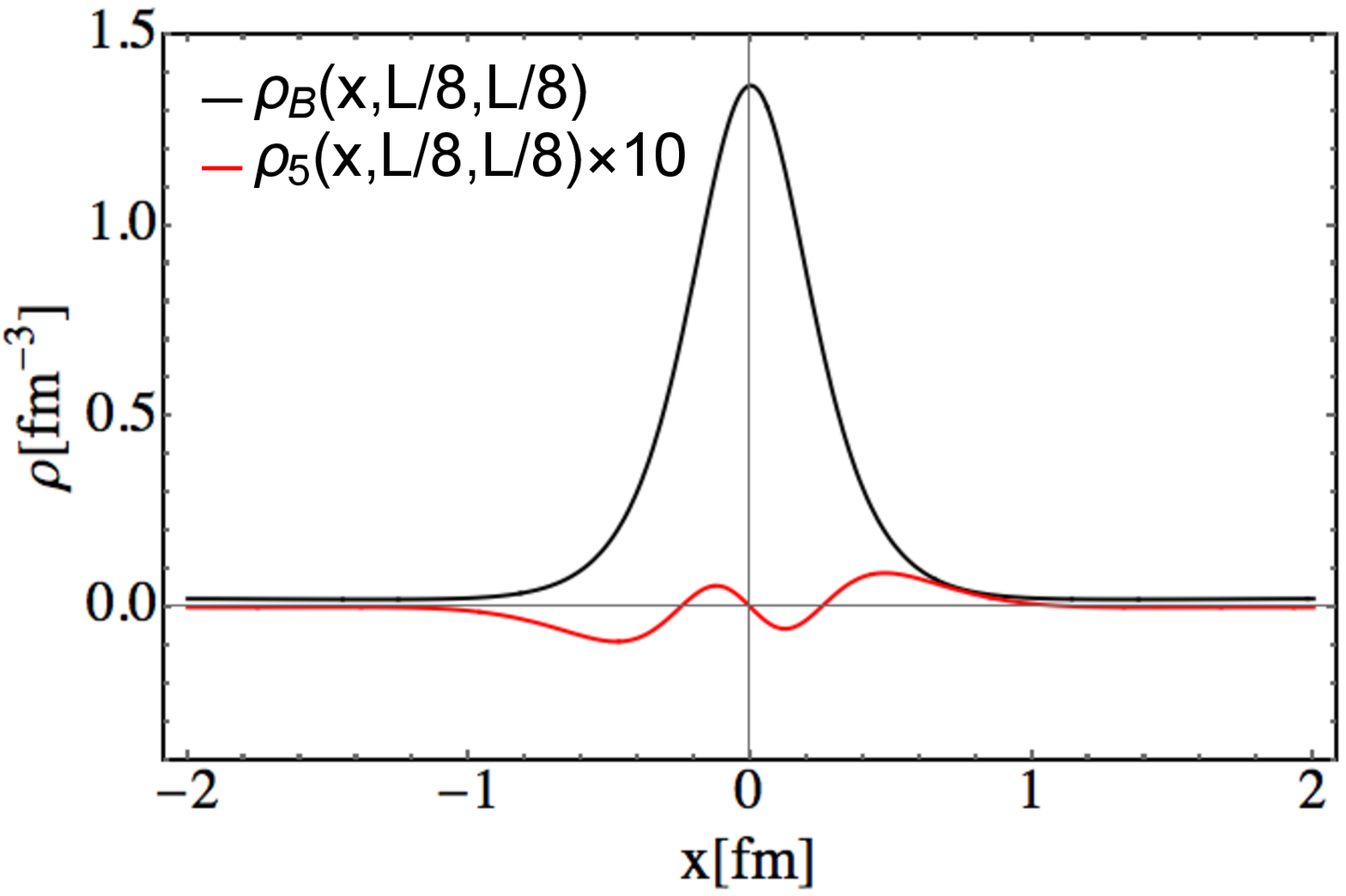}
    \subfigure{(c)}
  \end{center}
 \end{minipage}
 \end{tabular}
 \caption{ 
The chiral imbalance distribution on the skyrmion configuration at $L=2.0[{\rm fm}]$ for $\sqrt{eB}=400[{\rm MeV}]$.
(a) The contour plot of the skyrmion configuration on the x-y plane, $\rho_B(x, y, L/8)$.
(b) The contour plot of the chiral imbalance distribution on the x-y plane, $\rho_5(x, y, L/8)$. 
(c) The distribution of the skyrmion and the chiral imbalance along the x axis
specified at $y=z=L/8=0.25[{\rm fm}]$.}  
 \label{eB400L200}
\end{figure}

\begin{figure}[H]
\begin{tabular}{cc}
 \begin{minipage}{0.31\hsize}
  \begin{center}
   \includegraphics[width=5.5cm]{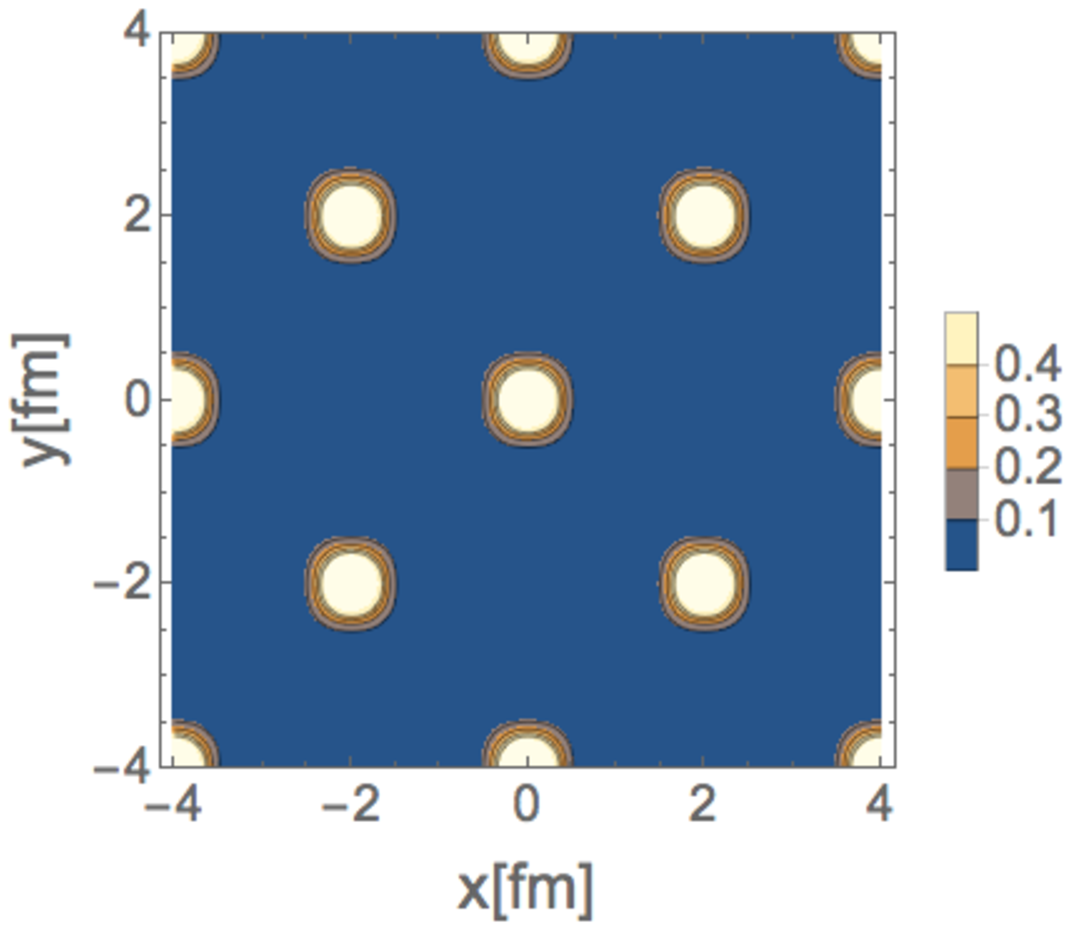}
    \subfigure{(a)}
  \end{center}
 \end{minipage} 
 \begin{minipage}{0.31\hsize}
  \begin{center}
   \includegraphics[width=5.5cm]{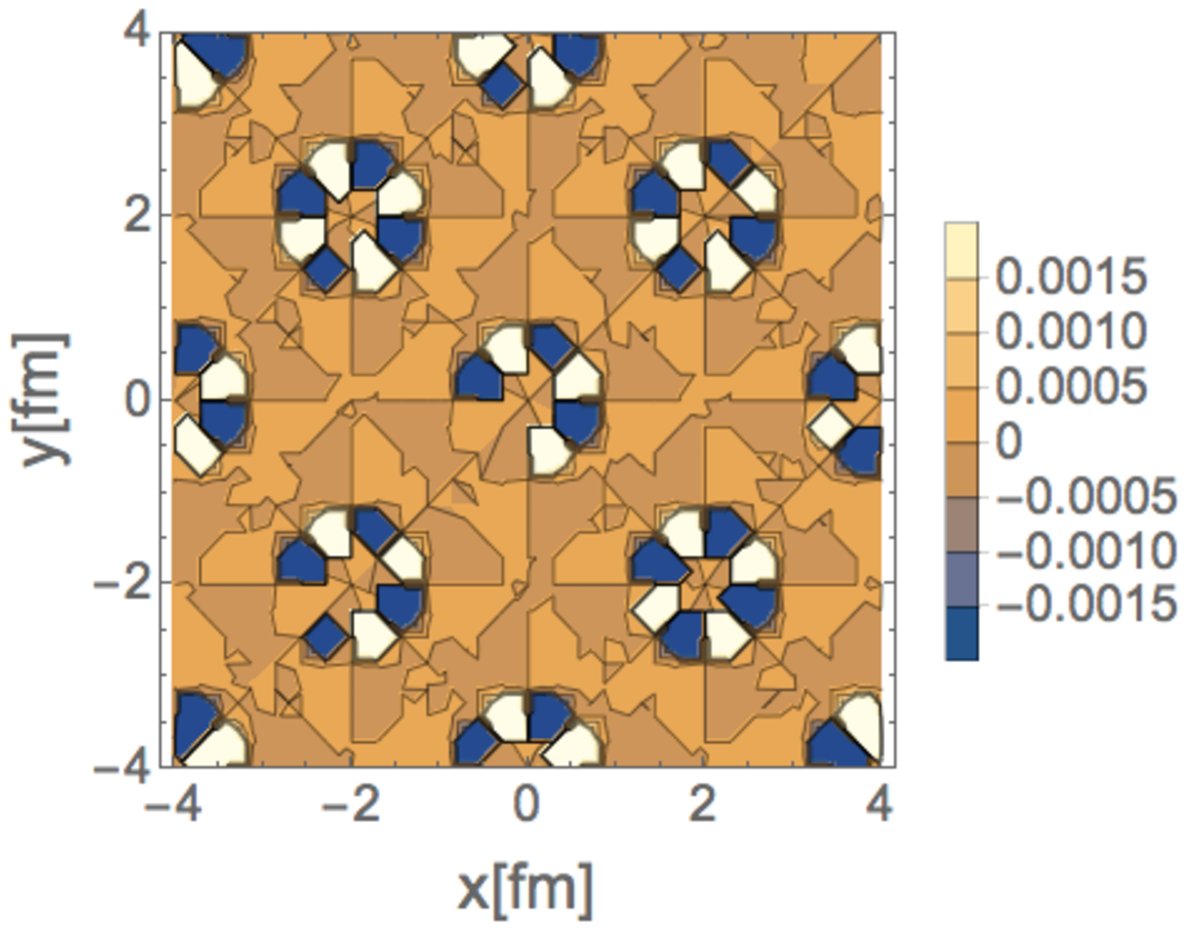}
    \subfigure{(b)}
  \end{center}
 \end{minipage}
  \begin{minipage}{0.31\hsize}
  \begin{center}
   \includegraphics[width=5.7cm]{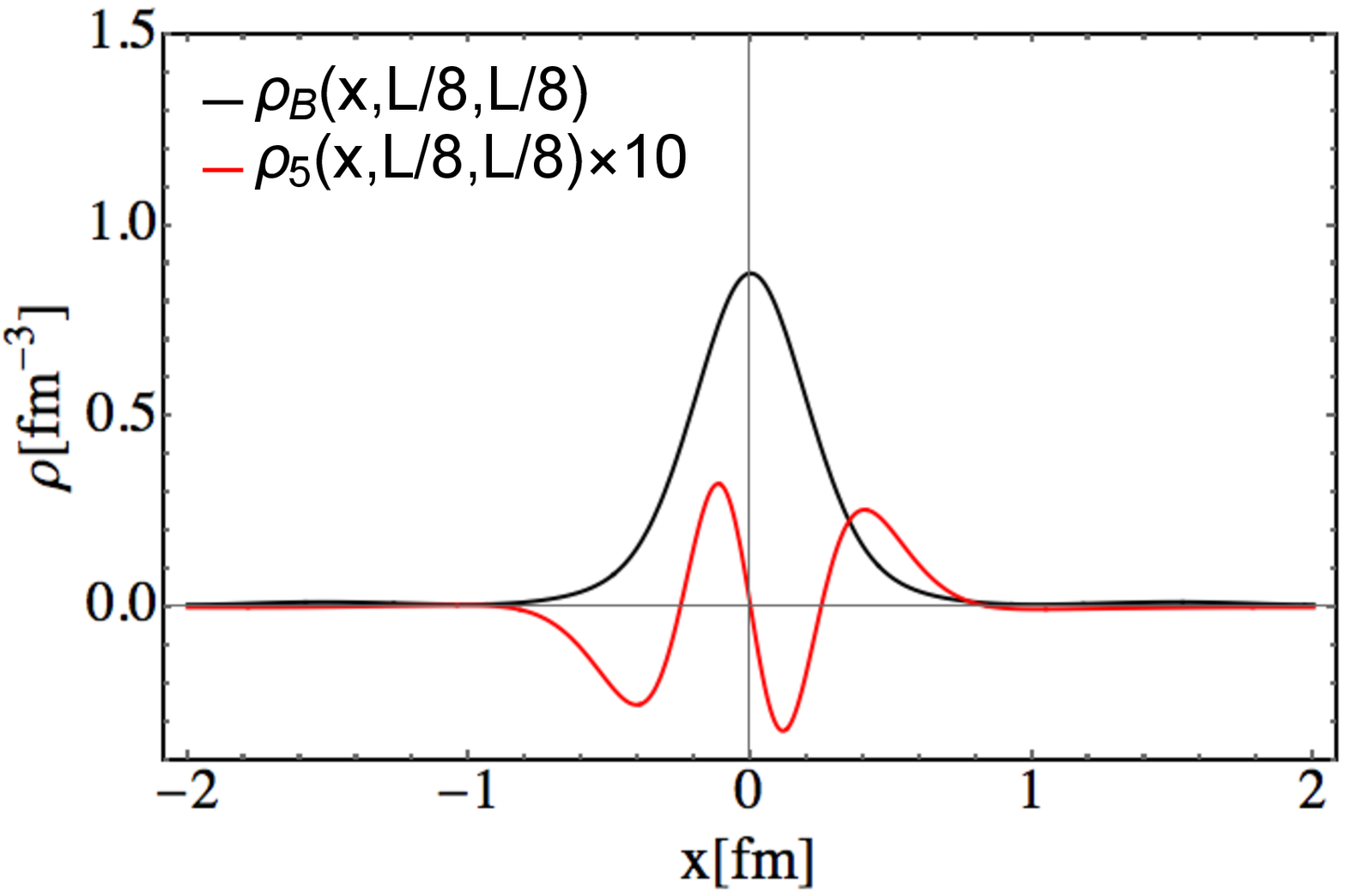}
    \subfigure{(c)}
  \end{center}
 \end{minipage}
 \end{tabular}
 \caption{ 
The chiral imbalance distribution on the skyrmion configuration at $L=2.0[{\rm fm}]$ for $\sqrt{eB}=800[{\rm MeV}]$.
(a) The contour plot of the skyrmion configuration on the x-y plane, $\rho_B(x, y, L/8)$.
(b) The contour plot of the chiral imbalance distribution on the x-y plane, $\rho_5(x, y, L/8)$. 
(c) The distribution of the skyrmion and the chiral imbalance along the x axis
specified at $y=z=L/8=0.25[{\rm fm}]$.}  
 \label{eB800L200}
\end{figure}

\begin{figure}[H]
\begin{tabular}{cc}
 \begin{minipage}{0.31\hsize}
  \begin{center}
   \includegraphics[width=5.5cm]{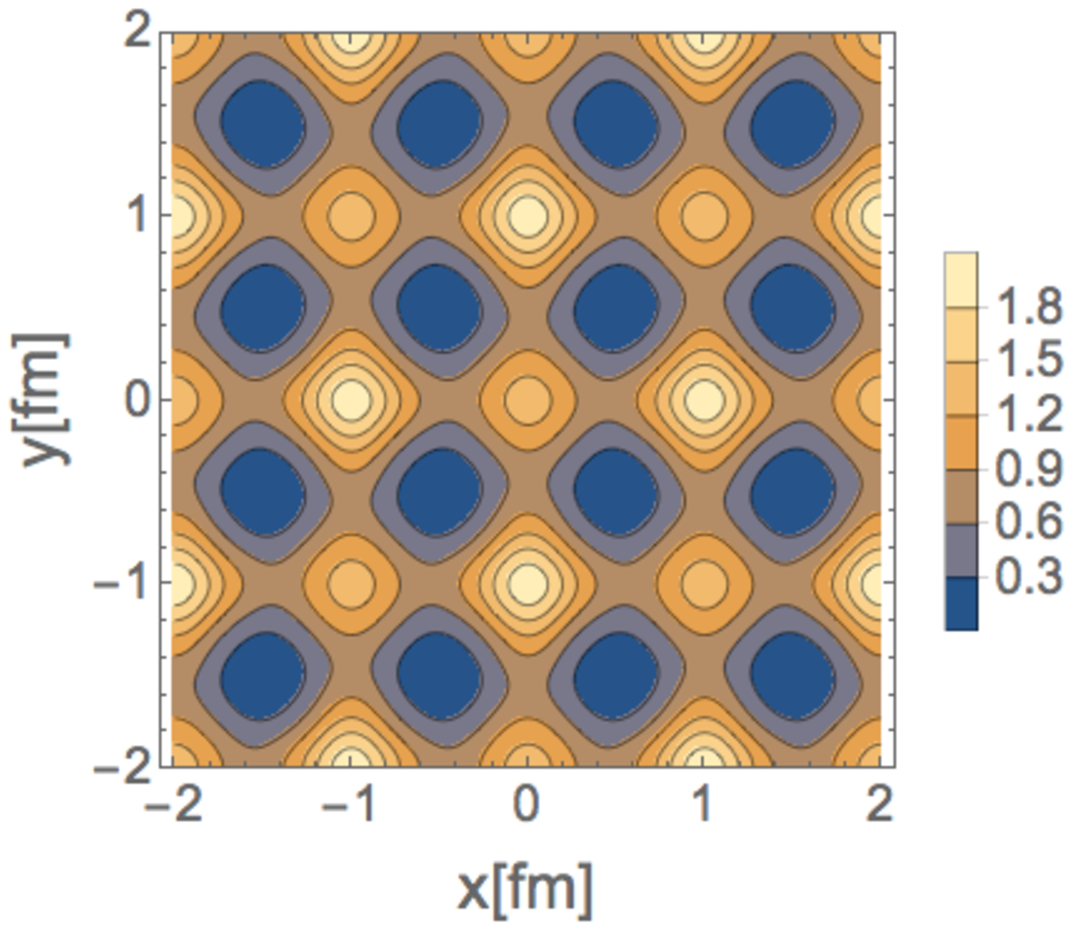}
    \subfigure{(a)}
  \end{center}
 \end{minipage} 
 \begin{minipage}{0.31\hsize}
  \begin{center}
   \includegraphics[width=5.5cm]{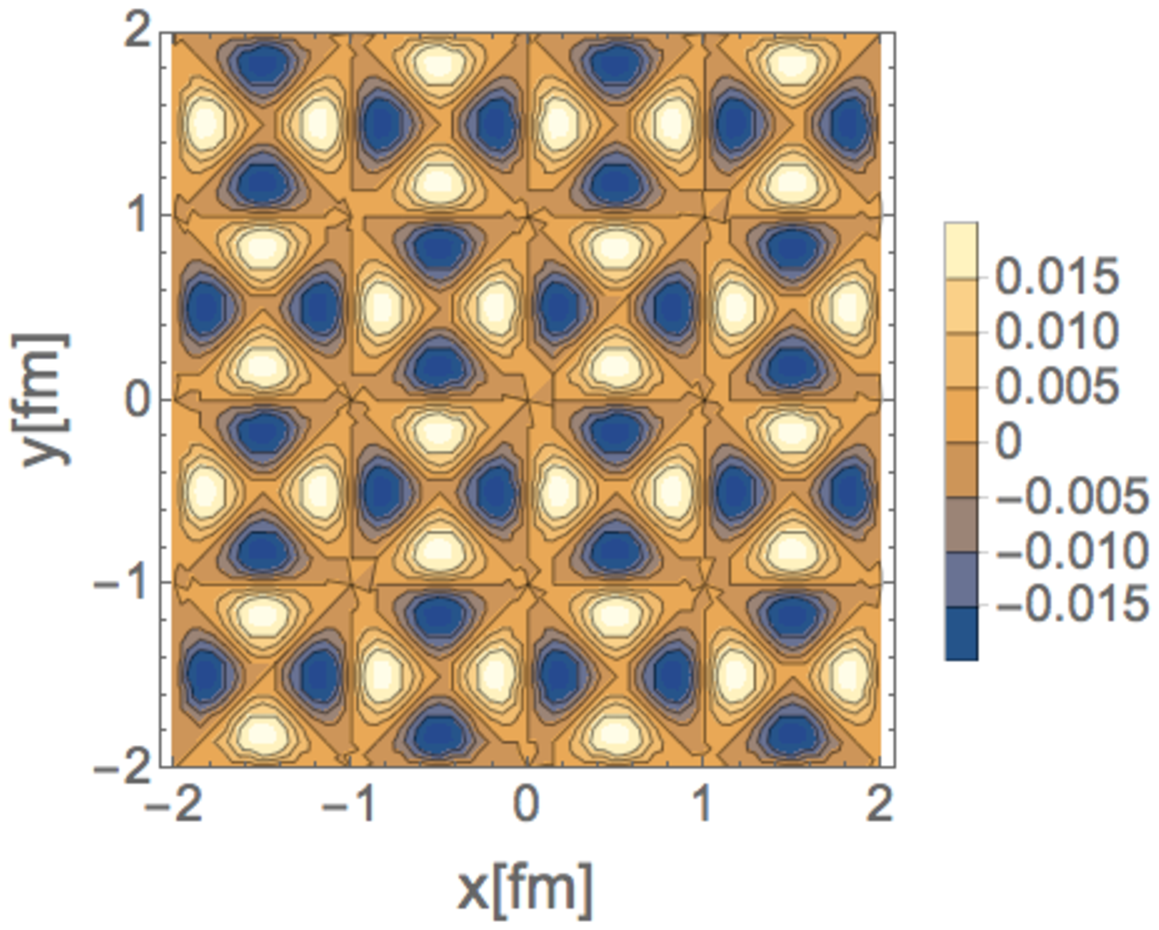}
    \subfigure{(b)}
  \end{center}
 \end{minipage}
  \begin{minipage}{0.31\hsize}
  \begin{center}
   \includegraphics[width=5.7cm]{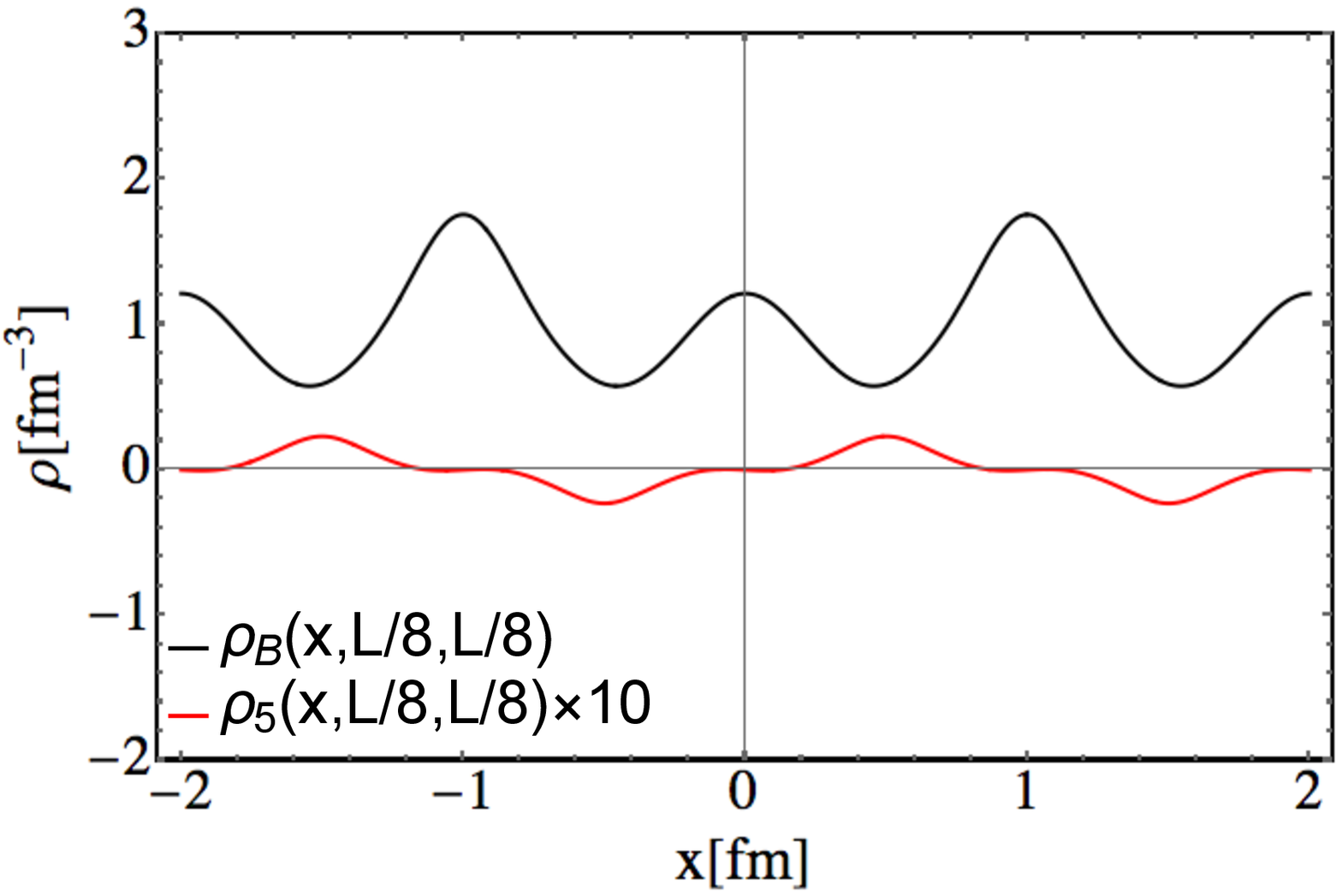}
    \subfigure{(c)}
  \end{center}
 \end{minipage}
 \end{tabular}
 \caption{ 
The chiral imbalance distribution on the skyrmion configuration at $L=1.0[{\rm fm}]$ for $\sqrt{eB}=400[{\rm MeV}]$.
(a) The contour plot of the skyrmion configuration on the x-y plane, $\rho_B(x, y, L/8)$.
(b) The contour plot of the chiral imbalance distribution on the x-y plane, $\rho_5(x, y, L/8)$. 
(c) The distribution of the skyrmion and the chiral imbalance along the x axis
specified at $y=z=L/8=0.125[{\rm fm}]$.}  
 \label{eB400L100}
\end{figure}

\begin{figure}[H]
\begin{tabular}{cc}
 \begin{minipage}{0.31\hsize}
  \begin{center}
   \includegraphics[width=5.5cm]{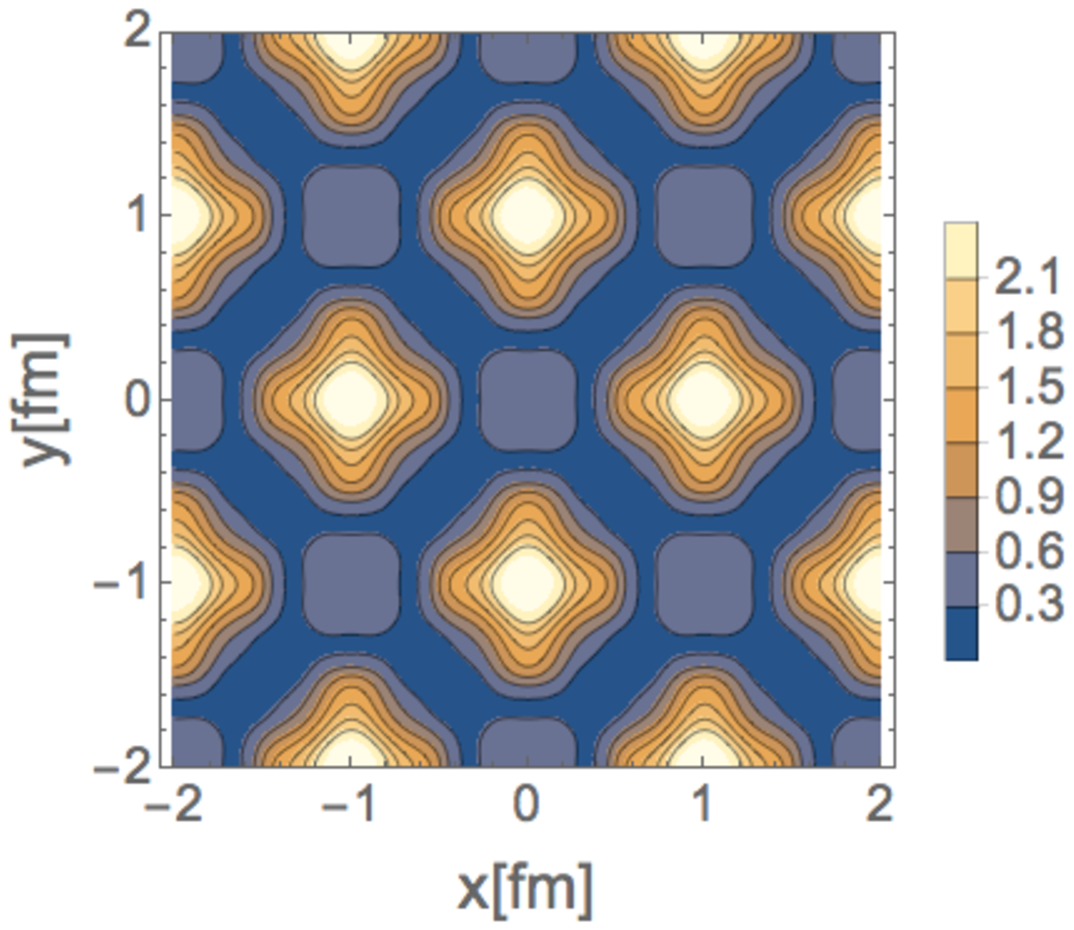}
    \subfigure{(a)}
  \end{center}
 \end{minipage} 
 \begin{minipage}{0.31\hsize}
  \begin{center}
   \includegraphics[width=5.5cm]{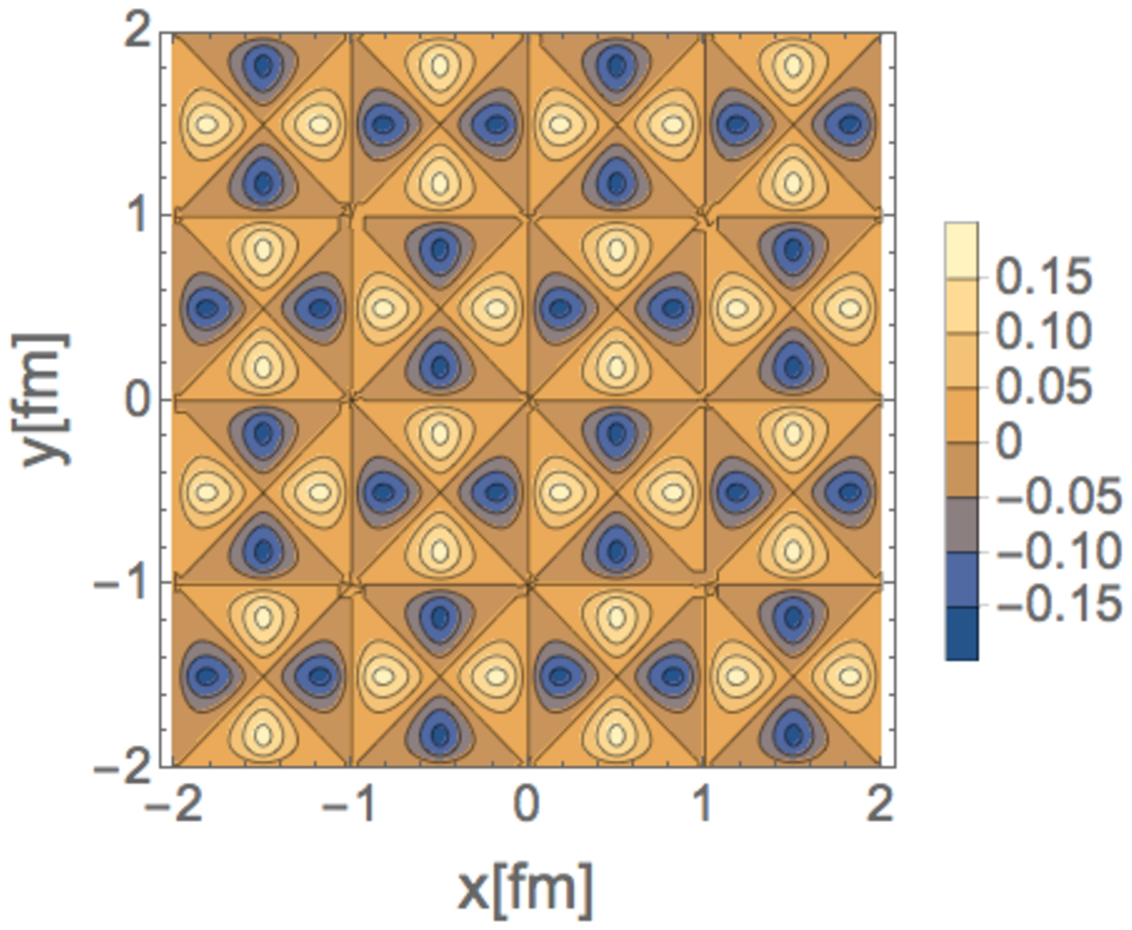}
    \subfigure{(b)}
  \end{center}
 \end{minipage}
  \begin{minipage}{0.31\hsize}
  \begin{center}
   \includegraphics[width=5.7cm]{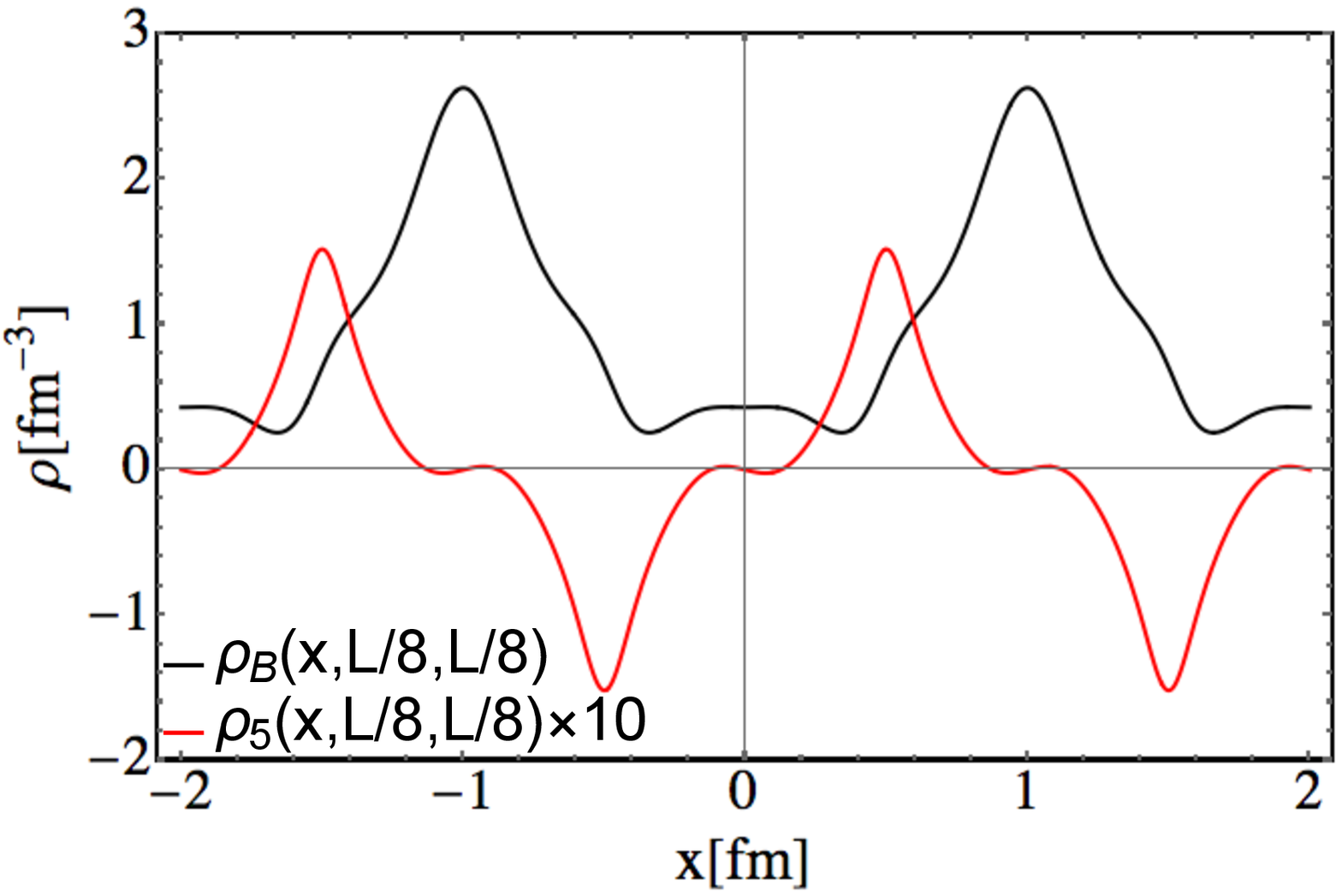}
    \subfigure{(c)}
  \end{center}
 \end{minipage}
 \end{tabular}
 \caption{ 
The chiral imbalance distribution on the skyrmion configuration at $L=1.0[{\rm fm}]$ for $\sqrt{eB}=800[{\rm MeV}]$.
(a) The contour plot of the skyrmion configuration on the x-y plane, $\rho_B(x, y, L/8)$.
(b) The contour plot of the chiral imbalance distribution on the x-y plane, $\rho_5(x, y, L/8)$. 
(c) The distribution of the skyrmion and the chiral imbalance along the x axis
specified at $y=z=L/8=0.125[{\rm fm}]$.}  
 \label{eB800L100}
\end{figure}

\end{widetext}

\subsection{Chiral-imbalance density wave and chiral density wave}  

In the previous work done by authors~\cite{Kawaguchi:2018xug}, 
a possible correlation between 
the inhomogeneous quark condensate and the deformation
of the skyrmion crystal form was addressed.  
In this subsection, 
we shall further show a remarkable presence of a nontrivial correlation 
between the chiral imbalance distribution, i.e., the chiral-imbalance density wave,  
and the inhomogeneous-chiral condensate distribution, i.e., a chiral density wave, 
as was observed in~\cite{Kawaguchi:2018xug}. 
As to the latter distribution, we may select the inhomogeneity of $\phi_1$ 
associated with the inhomogeneous quark condensate
(see Eq.(\ref{qcond})). 
This is because among the inhomogeneous chiral condensates $\sim \phi^\alpha$ 
in Eq.(\ref{qcond}), 
only the $\phi_1$ has the same parity property as the $\rho_5$ along the $x$-direction 
(see Eq.(\ref{ansatz_1})), hence it is most convenient to compare the $\rho_5$ distribution 
with that of the $\phi_1$ along the $x$-axis as has been depicted in the distribution 
plots so far.

Before proceeding the comparison between the $\rho_5$ and the $\phi_1$, 
we first note from Figs~\ref{eB400L200}, \ref{eB800L200}, \ref{eB400L100}, and \ref{eB800L100} that  
the magnitude of the chiral-inhomogeneous density wave is much smaller (by a factor of 10)
than the baryon number distribution (skyrmion configuration) in the system. 
Though it implies a small $U(1)_A$ anomaly effect, 
we may try to see how 
the local chirality imbalance is responsible for the 
nonzero inhomogeneity of $\phi_1$, 
in both the skyrmion and half-skyrmion phases,   
by comparing the wave forms and the peak structures for 
the chiral-inhomogeneous density wave and a chiral density wave. 
The observation like this would deduce a quantitative understanding 
on how much the $U(1)_A$ anomaly effect is encoded in the 
chiral condensate. 
To easily see and visually grasp a nontrivial correlation 
in such peak structures by eyes, 
we may amplify the amplitude of $\rho_5$ 
in the (half-) skyrmion phase, by a factor 
defined as 
\begin{align}
C_{400(800)}^{{\rm (h-)skyr}}
\equiv 
\frac{\phi_1^{\rm max}(\bar{x},y=z=L/8)|_{\sqrt{eB}=400(800)\, {\rm MeV}} } 
{\rho_5^{\rm max} (\bar{x}, y=z=L/8)|_{\sqrt{eB}=400(800) {\rm MeV}} } 
\,, 
\label{dimensionless}
\end{align}
where ``max'' denotes the maximum value realized at $(x, y, z)=(\bar{x}, L/8, L/8)$ 
for given $\sqrt{e B}$ 
(in which the $\bar{x}$ is just a number, to be read off).  
Thus, the amplified chiral-imbalance distribution, 
($\rho_5\times C_{400(800)}^{\rm (h-)skyr}$), is set to 
a dimensionless quantity as well as the $\phi_1$, and can have the same order of 
magnitude as what the $\phi_1$ can have.

In Fig.~\ref{L200phi1rho5} 
we plot the amplified chiral-imbalance distributions, ($\rho_5\times C_{400(800)}^{\rm skyr}$),
and the inhomogeneity distributions for $\phi_1$, 
in the skyrmion and half-skyrmion phases, respectively. 
From this figure, we find the following features: 
\begin{itemize} 

\item 
As the strength of a magnetic field increases, 
the inhomogeneity of $\phi_1$ tend to be localized and the amplitude of $\phi_1$ becomes small, as discussed in \cite{Kawaguchi:2018xug}.

\item 
As for the correlation between the chiral-imbalance density wave  
and a chiral density wave depicted by the inhomogeneity of $\phi_1$,
the peak point for the former 
does not 
match with that of the $\phi_1$-chiral density wave. 

\end{itemize}

Moving on to the half-skyrmion phase,
we make plots of the amplified chiral-imbalance distribution,
$\rho_5\times C_{400(800)}^{\rm h-skyr}$,
and the inhomogeneities of $\phi_1$ in Fig. \ref{L100phi1rho5}.
The figure tells us the following characteristic properties: 
\begin{itemize} 

\item 
In contrast to the skyrmion phase, 
the magnetic effect 
is insensitive to the inhomogeneous configuration of $\phi_1$,
as discussed in \cite{Kawaguchi:2018xug}.

\item 
The periodicity of the chiral-imbalance density wave 
($\rho_5\times C_{400(800)}^{\rm h-skyr}$) 
synchronizes with a chiral density wave formed by 
the $\phi_1$ inhomogeneity for any strength of a magnetic field.

\end{itemize}

It is of particular interest to note from the second item
that in the half-skyrmion phase
the presence of a magnetic field
makes a nontrivial correlation between the chirality imbalance and 
the inhomogeneous quark condense, 
in terms of the periodicity for density wave distributions. 
Actually, the coincidence of the periodicity in the half-skrymion phase 
can analytically be understood as 
by the symmetry properties for the $\rho_5$ and $\phi_1$ 
on crystals given 
in Eqs. (\ref{rho5trans:1}) - (\ref{phi1trans:2}), in Appendix A.

Those nontrivial-wave correlations having a different aspect between 
the skrymion and half-skyrmion phases 
would provide us with 
\cred{a novel possibility: 
the presence of the chiral-imbalance density wave 
as a consequence of the $U(1)_A$ anomaly}  
would be an important probe for the phase boundary between the 
skyrmion and half-skyrmion phases in the high-dense baryonic matter 
under a magnetic field.

\begin{widetext}

\begin{figure}[H]
\begin{tabular}{cc}
 \begin{minipage}{0.31\hsize}
  \begin{center}
   \includegraphics[width=5.5cm]{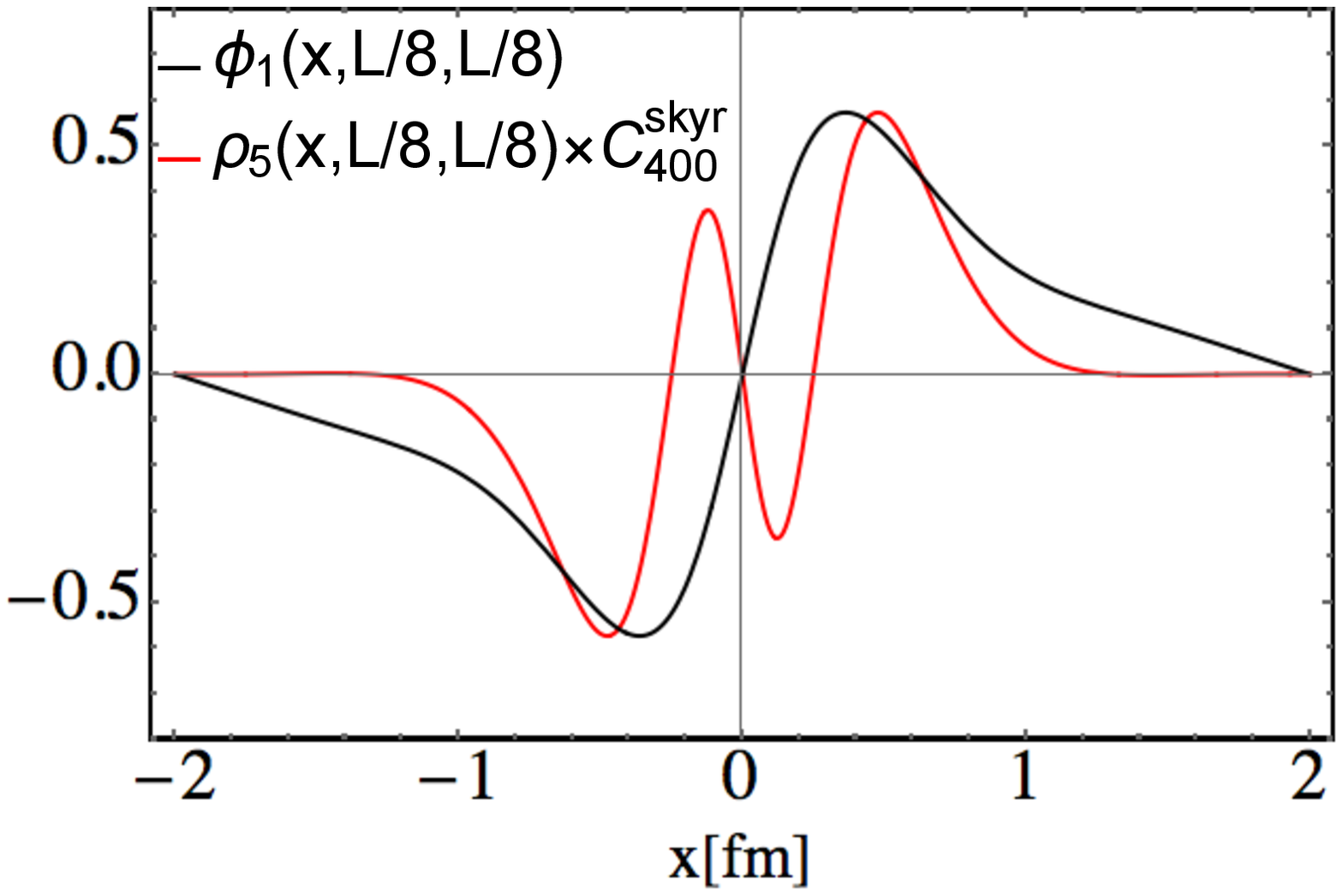}
    \subfigure{(a)}
  \end{center}
 \end{minipage} 
 \begin{minipage}{0.31\hsize}
  \begin{center}
   \includegraphics[width=5.5cm]{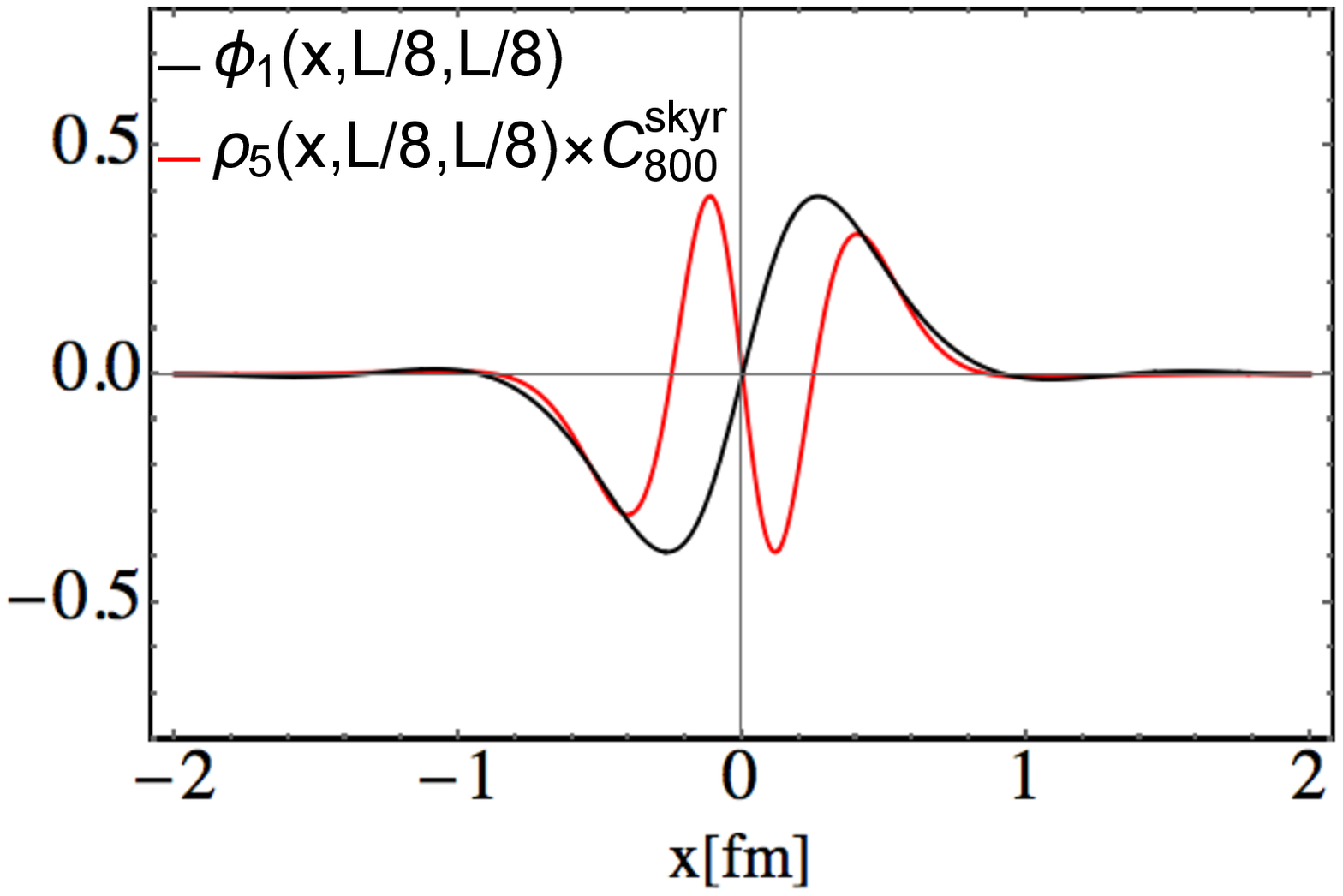}
    \subfigure{(b)}
  \end{center}
 \end{minipage}
 \end{tabular}
 \caption{ 
The distribution of $\phi_1(x,L/8,L/8)$ and $\rho_5(x, L/8, L/8)\times C_{400(800)}^{\rm skyr}$
in the skyrmion phase where
$L=2.0[{\rm fm}]$ for
$\sqrt{eB}=400[{\rm MeV}]$ (a) and 
$\sqrt{eB}=800[{\rm MeV}]$ (b). 
}  
 \label{L200phi1rho5}
\end{figure}

\begin{figure}[H]
\begin{tabular}{cc}
 \begin{minipage}{0.31\hsize}
  \begin{center}
   \includegraphics[width=5.5cm]{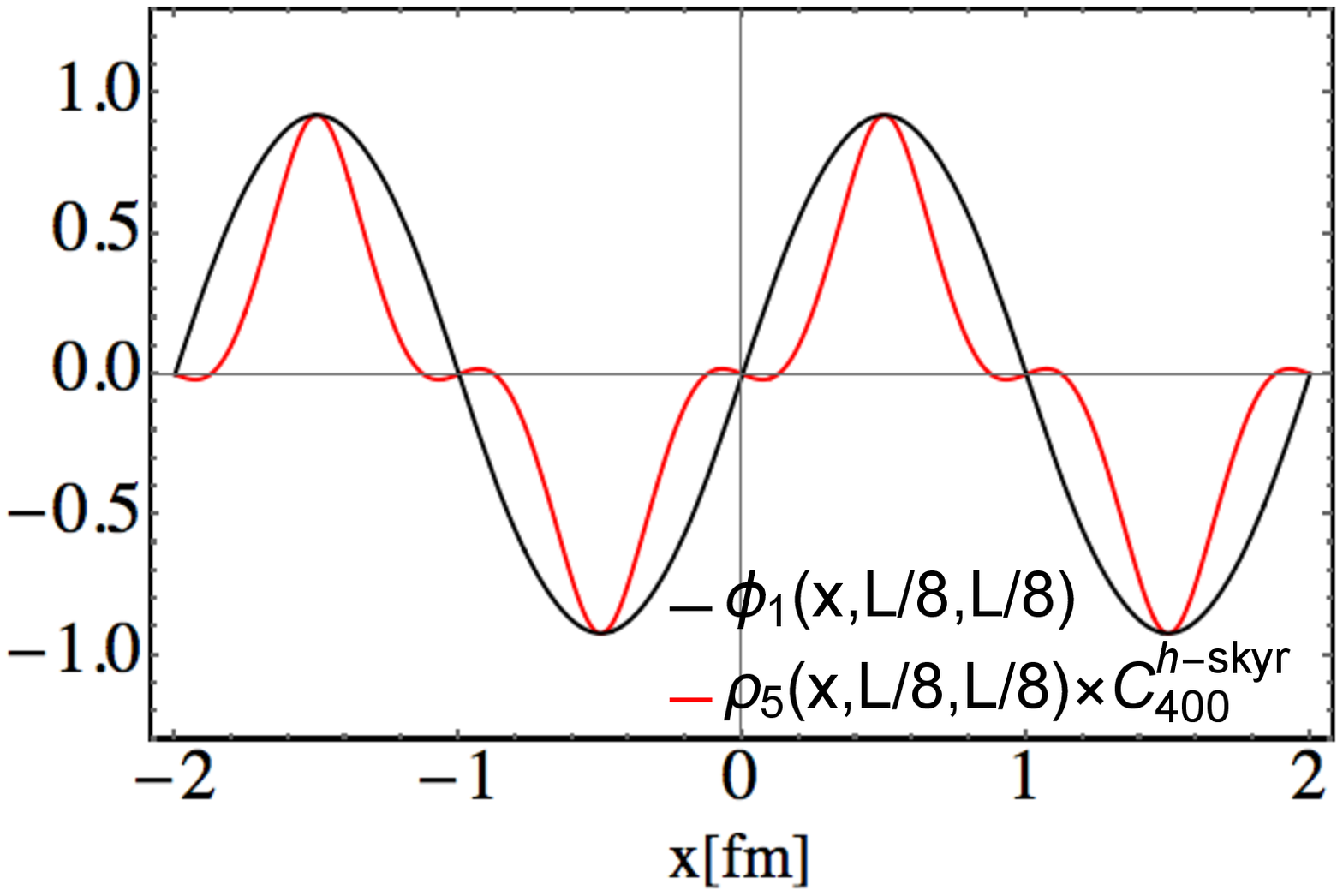}
    \subfigure{(a)}
  \end{center}
 \end{minipage} 
 \begin{minipage}{0.31\hsize}
  \begin{center}
   \includegraphics[width=5.5cm]{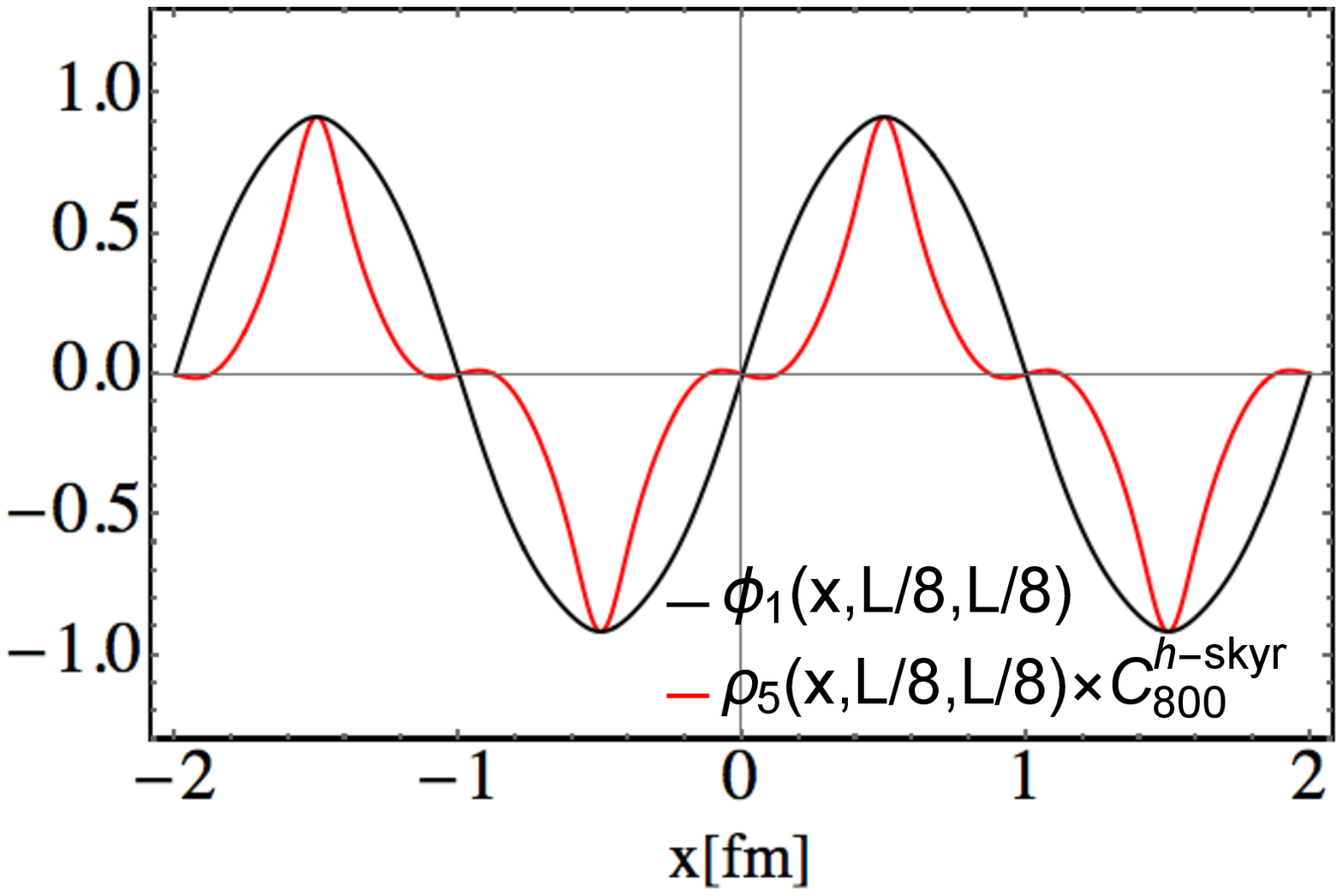}
    \subfigure{(b)}
  \end{center}
 \end{minipage}
 \end{tabular}
 \caption{ 
The distribution of $\phi_1(x,L/8,L/8)$ and $\rho_5(x, L/8, L/8)\times C_{400(800)}^{\rm (h-)skyr}$ 
in the half-skyrmion phase where $L=1.0[{\rm fm}]$ 
for $\sqrt{eB}=400[{\rm MeV}]$ (a) and 
 $\sqrt{eB}=800[{\rm MeV}]$ (b). 
}  
 \label{L100phi1rho5}
\end{figure}

\end{widetext}

\section{Conclusions}

In summary, 
we proposed a novel possibility to create a chiral-imbalance 
medium in a high dense baryonic matter under a 
magnetic field. 
It is a chirality-imbalance that can be emerged 
due to the magnetic $U(1)_A$ anomaly coupled 
with a local-nontrivial inhomogeneity of a 
pion-vector current arising 
in the high-dense matter system, as was 
roughly sketched in the introductory part 
of the present paper (Eq.(\ref{generic-rho5})). 
This imbalance is in contrast to the conventional one generated by 
the gluonic $U(1)_A$ anomaly in the case of 
hot QCD matter. 
Hence it would provide a new chance 
to examine how much the $U(1)_A$ anomaly can be 
relevant to the net chiral asymmetry, 
compared to the spontaneously broken 
chiral symmetry.

To demonstrate the crucial contribution of the proposed chiral-imbalance 
in Eq.(\ref{generic-rho5}), 
in the present paper  
we have taken the skyrmion crystal approach to make a model description  
for baryonic/high dense matters, to explicitly show that     
a nontrivial chiral imbalance distribution can indeed be 
induced in the modeled skyrmion crystal,  
due to the presence of a magnetic field. 
Interestingly enough, 
the chiral imbalance distribution turned out to take 
a wave form in a high density region, 
when the inhomogeneous chiral condensate 
develops to form a chiral density wave. 
This implies 
the emergence of 
a nontrivial density wave for the explicitly broken $U(1)_A$ current 
simultaneously with the chiral 
density wave for the spontaneously broken chiral-flavor current.  
This emergent wave was dobbed ``{\it chiral-imbalance density wave}''.

We further observed 
that the topological phase transition in the skyrmion crystal model 
(between the skyrmion and half-skyrmion phases) leads to  
the change of the chiral-imbalance density wave in shape.  
In particular, it was shown that 
in the half-skyrmion phase,  
the periodicity of the chirality-imbalance distribution
synchronizes with the the inhomogeneous chiral condensate, 
in contrast to the case of the skyrmion phase where 
the chiral-imbalance density wave flows with a different periodicity 
from a chiral density wave.

The emergence of the chiral-imbalance density wave in 
dense matters could give a crucial contribution 
to studies on the chiral phase transition,  
as well as the nuclear matter structure, 
in compact stars under a magnetic field, 
and would give a significant impact on analyses regarding 
the inhomogeneous chiral condensate through introducing 
the chiral density wave,  
as was mentioned in Introduction of the present paper. 

Also, our findings would make an important step to make deeper understanding 
of the role of the $U(1)_A$ anomaly in a sense of the origin of baryon mass 
as well as the baryon matter structure, 
and would give some impacts on an interdisciplinary physics 
like those raised in Introduction of this paper, 
e.g. the chirality imbalance for chiral neutrinos 
in supernova explosions and similar related chiral transport physics in 
dense matter systems.

\acknowledgments 

\cred{ 
We are grateful to Yong-Liang Ma for several useful comments.  
This work was supported in part by the JSPS Grant-in-Aid for Young Scientists (B) No. 15K17645 (S.M.), National Science Foundation of China (NSFC) under Grant No. 11747308 
(
S.M.), 
and the Seeds Funding of Jilin University (
S.M.). 
K.M. was also partially supported by the JSPS Grant-in-Aid for JSPS 
Research Fellow No. 18J1532. 
}

\appendix
\section{Translational symmetries for $\rho_5$ and $\phi_1$}
In this Appendix, we provide a supplement on  
the translational symmetry properties for $\rho_5$ and $\phi_1$, 
which can help understand the coincidence for the periodicity 
among them in the half-skyrmion phase, as has been 
observed in Fig.~\ref{L100phi1rho5}.

Under the translational symmetry,  $\rho_5$ transforms in the following way: 
in skyrmion phase with the FCC structure, we have 
\begin{align}
&\rho_5(x,y,z)=\rho_5(x+L,y+,L,z) 
\notag \\ 
& =-\rho_5(x+L,y,z+L)=-\rho_5(x,y+L,z+L)
\,, 
\label{rho5trans:1}
\end{align} 
while in half-skyrmion phase with the CC structure, we have 
\begin{align} 
&\rho_5(x,y,z)=-\rho_5(x+L,y,z) 
\notag \\ 
& =-\rho_5(x,y+L,z)
=\rho_5(x,y,z+L).
\label{rho5trans:2}
\end{align}

Under the translational symmetry,  $\phi_1$ transforms as follows: 
in skyrmion phase with the FCC structure, we have 
\begin{align} 
&\phi_1(x,y,z)=-\phi_1(x+L,y+L,z) 
\notag \\
& =-\phi_1(x+L,y,z+L)=\phi_1(x,y+L,z+L)\,, 
\label{phi1trans:1}
\end{align}
while in half-skyrmion phase with the CC structure, we have 
\begin{align}
&\phi_1(x,y,z)=-\phi_1(x+L,y,z)
\notag \\ 
& =\phi_1(x,y+L,z)=\phi_1(x,y,z+L) 
\,. 
\label{phi1trans:2}
\end{align}


\end{document}